\documentclass[12pt]{iopart}
\usepackage{graphicx}
\pdfoutput=1 

\begin{document}
\vspace*{-0.5cm}

\title[Nano-scale simulation of neuronal damage by galactic cosmic rays]{Nano-scale simulation of neuronal damage by galactic cosmic rays}

\author{Jonah S Peter$^1$, Jan Schuemann$^2$, Kathryn D Held$^2$, and Aimee McNamara$^2$}

\address{$^1$ Biophysics Program, Harvard University, Boston, MA 02115, United States of America}
\address{$^2$ Department of Radiation Oncology, Massachusetts General Hospital, Boston, MA 02114, United States of America}
\ead{jonahpeter@g.harvard.edu}
\vspace{10pt}
\begin{indented}
\item[]February 14, 2022
\end{indented}


\begin{abstract}
\underline{Objective:} 
The effects of complex, mixed-ion radiation fields on neuronal function remain largely unexplored. \textit{In silico} modeling studies of radiation-induced neuronal damage provide important quantitative information on physico-chemical processes that are not directly accessible through radiobiological experiments. Here, we present a complete analysis of the nano-scale physics associated with broad-spectrum galactic cosmic ray (GCR) irradiation in a realistic cornu ammonis 1 (CA1) pyramidal neuron geometry.

\underline{Approach:}
We simulate the entire 33 ion-energy beam fluence distribution currently in use at the NASA Space Radiation Laboratory galactic cosmic ray simulator (GCRSim). We use the TOol for PArticle Simulation (TOPAS) and TOPAS-nBio Monte Carlo-based track structure simulation toolkits to assess the dosimetry, physics processes, and fluence statistics of different neuronal compartments at the nanometer scale. We also make comparisons between the full GCRSim distribution and a simplified 6 ion-energy spectrum (SimGCRSim).

\underline{Main Results:} We find that the dominant dose contribution comes from electrons $(77 \pm 1\%)$, with smaller contributions coming from protons, $\alpha$-particles, and heavy ions. We show that across all physics processes, ionizations mediate the majority of the energy deposition $(68 \pm 1\%)$, though vibrational excitations are the most abundant ($70 \pm 2\%$ of all energy deposition events). We report that neuronal energy deposition by proton and $\alpha$-particle tracks declines approximately hyperbolically with increasing primary particle energy at mission-relevant energies. We also demonstrate an inverted exponential relationship between dendritic segment irradiation probability and neuronal absorbed dose. Finally, we find that there are no significant differences in the average physical responses between the GCRSim and SimGCRSim fluence distributions.

\underline{Significance:}
To our knowledge, this is the first nano-scale simulation study of a realistic neuron geometry using the GCRSim and SimGCRSim fluence distributions. The results presented here are expected to aid in the interpretation of future experimental results and help guide future study designs.
\end{abstract}

\vspace{2pc}

\section{Introduction}
As crewed spaceflight begins to return to the moon and extends towards Mars, astronauts will face significantly increased exposure to space radiation in the form of solar particle events (SPEs) and galactic cosmic rays (GCRs). While SPEs consist mainly of $< 10^2$ MeV protons, the GCR field contains significant dose contributions from high linear energy transfer (LET) ions ranging from hydrogen ($Z = 1$) to nickel ($Z = 28$) on the periodic table and energies ranging from $10 - 10^4$ MeV/n [1]–[5]. Although adequate shielding can significantly reduce the risk of SPE exposure, GCR radiation is less amenable to shielding and therefore poses a major challenge for long-duration spaceflight [6]. The development of accurate risk estimation tools and alternative countermeasures remains a major focus of active research, but is complicated by large uncertainties in characterizing the radiobiological responses of living tissues. Difficulties stem, in part, from the diverse nature of tissue-specific radiation-induced sequelae as well as the technical challenges of emulating the complex GCR field in a laboratory setting.

Of particular concern for both mission success and long-term astronaut health is sustained radiation exposure to the central nervous system (CNS). There now exists a plethora of research indicating that single-particle exposures can result in significant cognitive deficits at mission-relevant doses and energies. Studies in mice and rats have demonstrated both short-term ($1-2$ weeks) and persistent (1 year) behavioral changes post-irradiation across a variety of particle species including neutrons, protons, $\alpha$-particles, $^{12}$C, $^{16}$O, $^{28}$Si, $^{48}$Ti, and $^{56}$Fe at doses as little as 50 mGy [7]. These studies indicate deficits in attention [8], spatial learning [9]–[15], object recognition [10], [13], [15]–[17], episodic memory [10], [13], fear conditioning [9], [10], [13], [15], [16], [18], locomotor activity [19], [20], social behavior [10], and anxiety responses [9], [10], [13], [15], [18], among other behaviors. Additionally, clinical studies in patients treated with radiation therapy have shown impairments to cognition and memory, demonstrating that these deleterious effects also manifest in humans, albeit at higher doses [21]–[24]. Significant deficits to cognitive function during spaceflight could pose a serious threat to astronaut safety and overall mission success.

One brain structure which has remained the focus of extensive radiobiological interest is the hippocampus. Learning and memory are influenced by the modulation of glutamatergic synapses in the hippocampal cornu ammonis 1 (CA1) region leading to long-term potentiation (LTP). Dendritic spines on postsynaptic CA1 pyramidal neurons containing N-methyl-D-aspartate (NMDA) and $\alpha$-Amino-3-hydroxy-5-methyl-4-isoxazolepropionic acid (AMPA) receptors mediate the strengthening of these synapses in response to repeated activation [25]. As such, radiation-induced damage to the dendrites, dendritic spines, and synaptic receptors of CA1 pyramidal neurons is expected to cause widespread cognitive decline. Morphological studies have reported post-irradiation changes in dendritic length, branching, complexity, spine density, and spine type [12], [15], [19], [26]. Electrophysiology studies have demonstrated reductions in functional connectivity, neuronal excitability, and LTP responses several months post-irradiation—indicating long-term reductions in synaptic plasticity [9]–[11], [15], [27]. Molecular analyses have also demonstrated radiation-induced reductions in the glutamate readily releasable pool and postsynaptic NMDA receptor expression in the CA1 synapse [28]. These molecular changes have been linked to behavioral deficits which persist for at least 6 months [28]. 

However, despite the overwhelming evidence for GCR-induced CNS damage, accurate risk estimation for astronauts remains challenging. In particular, most radiobiology experiments have examined only a limited number of ion-energy combinations. Only in the last few years has it become possible to experimentally assess the neurobiological effects of chronic exposure to a broad-spectrum, mixed-ion radiation field akin to the natural GCR environment through ground-based studies. In 2018, researchers at the NASA Space Radiation Laboratory at Brookhaven National Laboratory performed the first radiobiological experiments using the newly-developed GCR simulator (GCRSim) designed to emulate the GCR exposure to blood-forming organs during realistic spaceflight conditions. The simulated spectrum consists of 33 different ion-energy combinations and closely resembles the dose distribution produced within the human body after 20 g cm$^{-2}$ of aluminum shielding (table 1) [3]. The facility also provides access to a simplified 6 ion-energy beam distribution (SimGCRSim) for smaller-scale studies (table 2). Experiments to examine CNS dysfunction, cognitive impairment, Alzheimer’s pathology, and to test preventative biological countermeasures are now underway [3]. 
\begin{table}
\caption{\label{1} Fluence distribution for the GCRSim spectrum. Values are normalized to a total dose of 0.5 Gy. In this work, the 1000 MeV/n $^{48}$Ti fluence was replaced with 1000 MeV/n $^{56}$Fe as discussed in the text. (p: proton; $\alpha$: $\alpha$-particle). All values obtained from Simonsen et al. [3].}
\lineup
\footnotesize
\begin{indented}
\item[]\begin{tabular}{@{}lll}
\br
Ion & Energy (MeV/n) & Dose (Gy)\\
\mr
p & \0\020 & 0.0304\\
p & \0\023 & 0.0067\\
p & \0\027 & 0.0074\\
p & \0\032 & 0.008\\
p & \0\037 & 0.0087\\
p & \0\043 & 0.0093\\
p & \0\050 & 0.01\\
p & \0\059 & 0.0106\\
p & \0\069 & 0.0111\\
p & \0\080 & 0.0112\\
p & \0100 & 0.0272\\
p & \0150 & 0.035\\
p & \0250 & 0.0689\\
p & 1000 & 0.1236\\
$\alpha$ & \0\020 & 0.011\\
$\alpha$ & \0\023 & 0.0021\\
$\alpha$ & \0\027 & 0.0022\\
$\alpha$ & \0\032 & 0.0023\\
$\alpha$ & \0\037 & 0.0025\\
$\alpha$ & \0\043 & 0.0026\\
$\alpha$ & \0\050 & 0.0027\\
$\alpha$ & \0\059 & 0.0027\\
$\alpha$ & \0\069 & 0.0027\\
$\alpha$ & \0\080 & 0.0027\\
$\alpha$ & \0100 & 0.00061\\
$\alpha$ & \0150 & 0.0075\\
$\alpha$ & \0250 & 0.0164\\
$\alpha$ & 1000 & 0.0249\\
$^{12}$C & 1000 & 0.0117\\
$^{16}$O & \0350 & 0.0154\\
$^{28}$Si & \0600 & 0.0081\\
$^{56}$Fe & \0600 & 0.0041\\
$^{48}$Ti ($^{56}$Fe) & 1000 & 0.0045\\
\br
\end{tabular}
\end{indented}
\end{table}

\begin{table}
\caption{\label{2} Fluence distribution for the SimGCRSim spectrum. Values are normalized to a total dose of 0.5 Gy. (p: proton; $\alpha$: $\alpha$-particle). All values obtained from Simonsen et al. [3].}
\lineup
\footnotesize
\begin{indented}
\item[]\begin{tabular}{@{}lll}
\br
Ion & Energy (MeV/n) & Dose (Gy)\\
\mr
p & \0250 & 0.0195\\
p & 1000 & 0.175\\
$\alpha$ & \0250 & 0.09\\
$^{16}$O & \0350 & 0.03\\
$^{28}$Si & \0600 & 0.005\\
$^{56}$Fe & \0600 & 0.005\\
\br
\end{tabular}
\end{indented}
\end{table}
With a plethora of new data on the horizon, additional \textit{in silico} modeling studies will be necessary to accompany the advancements in experimental design. Quantification of physico-chemical parameters that are not directly accessible through experiments are crucial for linking the earliest stages of radiation damage to concomitant biological responses and eventual physiological outcomes. Additionally, nanometer-scale simulation studies may further offer new insights into the most effective avenues of intervention for countermeasures. Although several studies have previously performed nano-scale physics simulations of radiation damage in the brain, none have addressed the more complex fluence distributions currently being used in experiments [29]–[32]. In the present work, we perform the first nanometer-scale modeling study of the full GCRSim fluence distribution in a realistic neuron geometry. We provide a complete analysis of the low-energy physics processes and particle species which mediate energy deposition. Comparisons are also made to the SimGCRSim distribution. The results presented here are expected to aid in the interpretation of future experimental results and can help guide future study designs.

\section{Methods}

\subsection{CA1 Pyramidal Neuron Geometry}

Morphological data for CA1 pyramidal neurons were obtained from adult ($2-3$ month old) wildtype C57Bl/6J mice from the NeuroMorpho.org online repository [33]. The age range and strain were selected for their ubiquitous use in radiobiology experiments. At the time of data collection there were 13 such neuron geometries available on the repository with complete information on dendrites and somas, all from the Soltesz archive (NeuroMorpho IDs NMO\_36611 through NMO\_36623) [34]. A single representative neuron (NMO\_36611) was selected with cumulative dendritic length, total surface area, and total volume of 4391.4 $\mu$m, 19821.6 $\mu$m$^2$, and 7316.8 $\mu$m$^3$, respectively. The measurements for this neuron are within one standard deviation of the mean values for the archive ($5236.3 \pm 1063.8$ $\mu$m, $19585 \pm 5463.3$ $\mu$m$^2$, $6248.9 \pm 2405.3$ $\mu$m$^3$ respectively; see supplemental table S1). The soma and dendrites were constructed \textit{in silico} according to the specifications in the NeuroMorpho parameter file. The soma was modeled as a Boolean intersection of two spheres, while the apical and basal dendritic trees were constructed from 1395 individual right circular cylinders (865 apical, 530 basal) with varying lengths ($3.15 \pm 1.37$ $\mu$m), radii ($0.712 \pm 0.120$ $\mu$m), and orientations (figure 1). The total \textit{in silico} volumes were 1166.0 $\mu$m$^3$ and 7220.5 $\mu$m$^3$ for the soma and dendrites, respectively.

\begin{figure}
    \includegraphics[width=\textwidth]{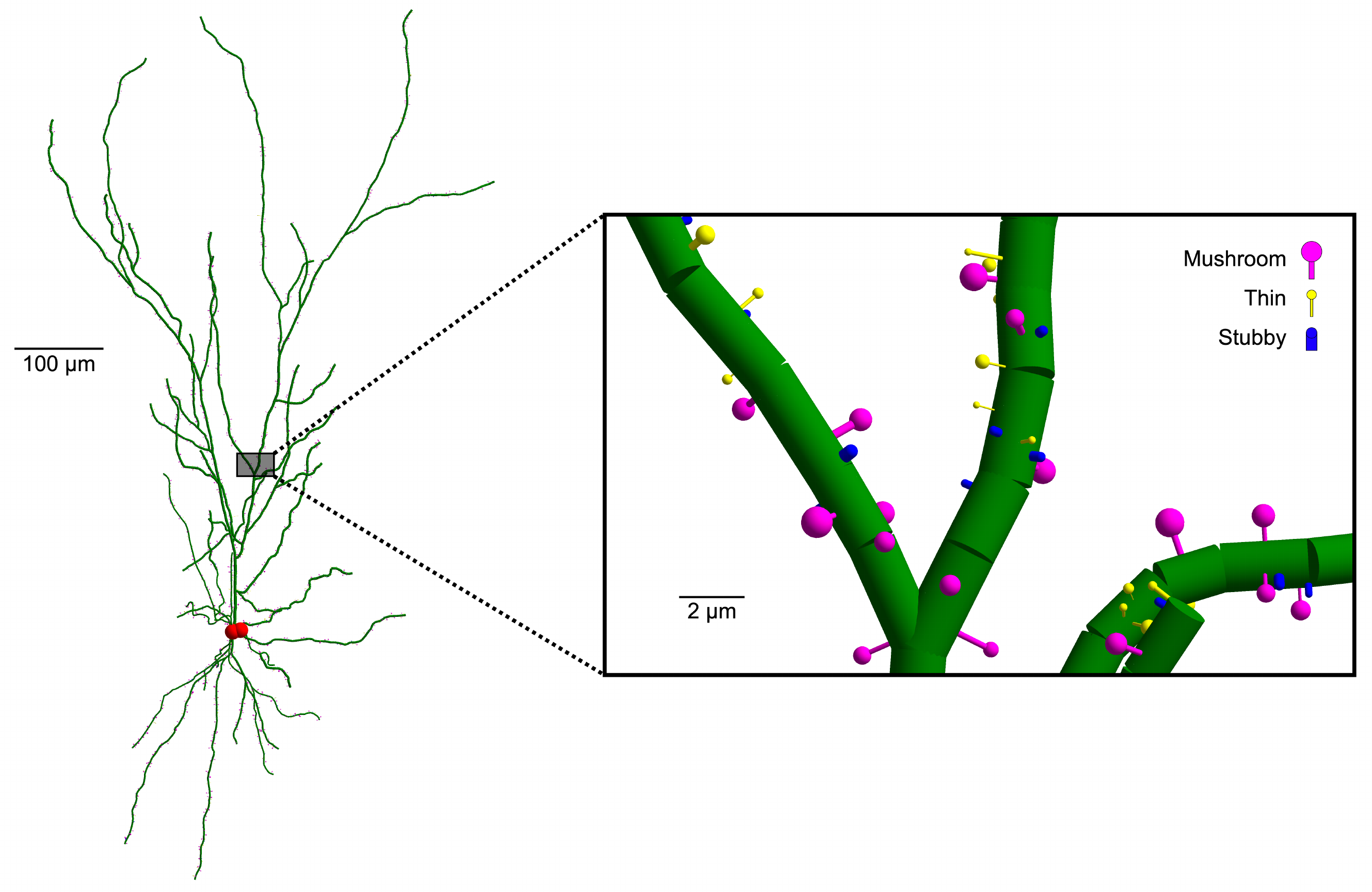}
    \caption{\textit{In silico} reconstruction of the neuron geometry used in this work. Apical and basal dendrites (green) are comprised from right circular cylinders, while the soma (red) is modeled as a Boolean intersection of spheres. Sizes, positions, and orientations of each volume compartment were obtained from NeuroMorpho (NMO\_36611 [34]) and constructed \textit{in silico} using TOPAS-nBio. \textbf{(Inset)} Mushroom (pink), thin (yellow), and stubby (blue) spines constructed from Boolean intersections of right circular cylinder stalks and spherical heads. Spine parameters were generated stochastically based on experimental data as discussed in the text.}
    \label{fig:1}
\end{figure}

\subsection{Stochastic Construction of Dendritic Spines}

Neuron geometries from the NeuroMorpho repository do not contain information about dendritic spines. Instead, each spine was constructed from a Boolean intersection of a spherical “head” centered atop a right circular cylinder “stalk” based on experimental measurements taken from the literature (figure 1, inset). Each spine was generated as one of three types: mushroom, thin, or stubby. For each mushroom and thin spine, the total spine length, head radius, and stalk radius were drawn stochastically from log-normal distributions with means and standard deviations obtained from the literature (table 3) [35], [36]. For stubby spines, which have no discernable distinction between head and stalk, the total length and head radius of each spine were generated stochastically while the stalk radius was set equal to the head radius. The relative fractions of spine types and overall spine densities were set separately for the apical and basal dendrites based on experimental data from the literature (table 3) [20], [26]. Spines were placed along the entire length of each dendrite without overlap and set orthogonal to the local longitudinal dendritic axis. The locations and azimuthal orientations of each spine were selected randomly according to uniform distributions. New spine populations were generated independently for each simulation run described below. While the relative fractions of spine types were drawn randomly from uniform distributions, the overall number of spines was kept fixed between runs. The total \textit{in silico} volume of each spine population varied as 246.6 $\pm$ 143.6 $\mu$m$^3$ across each run.

\begin{table}
\caption{\label{3} Parameters used to construct dendritic spine populations for each simulation run. For mushroom and thin spines, the length, stalk radius, and head radius of each spine were drawn stochastically from log-normal distributions. For stubby spines, the length and head radius of each spine were drawn stochastically and the stalk radius was set equal to the head radius. The number of apical and basal spines per micron of dendrite was held fixed between simulations at 0.8 and 0.85 $\mu$m$^{-1}$, respectively. The relative spine fractions were allowed to vary according to uniform distributions. Values are mean $\pm$ standard deviation across all runs.}
\lineup
\footnotesize
\begin{tabular}{@{}lllllll}
\br
&&&&&\centre{2}{Total Number$^{\ddagger}$}\\
\ns
\ns
&Length &Stalk Radius &Head Radius &Volume & \crule{2}\\
Spine Type & ($\mu$m)$^{\dagger}$ & ($\mu$m)$^{\dagger}$ & ($\mu$m)$^{\dagger}$ & ($\mu$m$^3$) & Apical & Basal\\
\mr
Mushroom & $1.50 \pm 0.25$ & $0.100 \pm 0.035$ & $0.400 \pm 0.064$ & $0.169 \pm 0.067$ & $\0740 \pm 210$ & $370 \pm 110$\\
Thin & $0.98 \pm 0.42$ & $0.050 \pm 0.015$ & $0.19\0 \pm 0.32$ & $0.021 \pm 0.070$ & $1010 \pm 290$ & $600 \pm 170$\\
Stubby & $0.44 \pm 0.15$ & --- & $0.160 \pm 0.065$ & $0.031 \pm 0.026$ & $\0550 \pm 160$ & $310 \pm 90$\\
\br
\end{tabular}\\
\noindent $^{\dagger}$Denotes values adapted from Harris et al. [35].\\
\noindent $^{\ddagger}$Denotes values adapted from Chakraborti et al. [26] and Allen et al. [20].
\end{table}

\subsection{Track Structure Simulations}

As high-energy particles interact with matter, they produce tracks of secondary particles which themselves interact with the irradiated material. The structure of these tracks generally depends upon the incident particle species and energy as well as the chemical composition of the material. Monte Carlo-based particle transport codes can accurately reproduce experimentally observed track structures and are widely utilized for biomedical and space-based applications. In this study, Monte Carlo simulations were performed using the TOol for PArticle Simulation (TOPAS) platform (version 3.4) originally developed for use in proton therapy [37], [38]. TOPAS wraps and extends the GEometry ANd Tracking 4 (Geant4) simulation toolkit [39]–[41]. To model particle track structure, we used the TOPAS-nBio package [42] which further extends TOPAS to radiobiological applications by providing geometry classes for cellular and sub-cellular structures. TOPAS-nBio includes a specific interface which reads and imports NeuroMorpho SWC files [43]. The neuron geometry used in this study was implemented through custom-built geometry and scoring classes using version 1.0-beta of TOPAS-nBio.

Tracks were simulated using the Geant4-DNA physics processes (Geant4 version 10-05-patch-01) which were implemented through TOPAS-nBio using the TsEmDNAPhysics physics list with default settings provided by TOPAS-nBio [44]–[48]. Accurate quantification of cellular (micron) and sub-cellular (nanometer) scale dosimetric quantities requires simulating electromagnetic interactions down to very low energies. The detailed physics models associated with the different Geant4-DNA processes are discussed elsewhere [44]–[47]. Briefly, nanometer-scale track structure is modeled by Geant4-DNA using a step-by-step single-scattering Monte Carlo approach. Cross sections for electromagnetic interactions with liquid water are available for electrons up to 1 MeV, protons, $\alpha$-particles, neutral $^1$H atoms, neutral $^4$He atoms, and singly ionized $^4$He atoms up to 100 MeV/n, and $^7$Li, $^9$Be, $^{11}$B, $^{12}$C, $^{14}$N, $^{16}$O, $^{28}$Si, and $^{56}$Fe ions up to 1 GeV/n. $\gamma$-radiation cross sections are also included up to 100 TeV based on the Geant4 Livermore models. The TsEmDNAPhysics list provided by TOPAS extends proton cross sections to 500 MeV and selects optimized physics models to match experimentally verified yields of common radiolysis products [48]. Energy deposition events including ionizations, electronic excitations, vibrational excitations, and electron attachments are tracked for particle energies down to 7.4 eV. For electrons with energies below this cutoff, the remaining energy is deposited locally and the electrons are considered solvated. Additional hadronic processes were simulated using the g4h-phy\_QGSP\_BIC\_HP, g4decay, g4ion-binarycascade, g4h-elastic\_HP, and g4stopping physics lists provided by TOPAS.

The simulation geometry was a 404 $\mu$m $\times$ 904 $\mu m$ $\times$ 104 $\mu m$ box with the neuron geometry placed at the center. All volumes were filled with 1 g cm$^3$ density water (Geant4 material G4\_WATER). Primary particles were randomly initialized on the surface of the box with initial momentum vectors oriented randomly towards the interior in order to simulate an isotropic flux. The fluence spectrum was chosen to match the ions, energies, and dose fractions used in GCRSim and SimGCRSim experiments at the NASA Space Radiation Laboratory in Brookhaven (tables 1 and 2). However, since $^{48}$Ti cross sections are not supported by Geant4-DNA, the 1000 MeV/n $^{48}$Ti fluence was substituted with 1000 MeV/n $^{56}$Fe as a conservative estimate.

\subsection{Calculation of Dosimetric and Fluence Quantities}

All simulations were performed on the ERISONE Linux cluster at Massachusetts General Hospital. For each ion-energy combination in the GCRSim spectrum (entries in table 1), $2\times 10^4$ primary particles and all of their secondaries were simulated. Computations were parallelized by splitting each ion-energy simulation into 100 independent runs of 200 primary particles each. Energy deposition events for each run were scored by subcellular region, physics process, and mediating particle species. Runs were pooled into groups of 10 to produce 10 pooled runs of $2\times10^3$ primary particles each. Scored quantities were then averaged across the pooled runs and standard deviations were computed to estimate their variability. As new populations of stochastically generated spines were constructed independently for each run, standard deviations incorporate variability in both the irradiation process and also CA1 pyramidal neuron morphology. 

Since proton and $\alpha$-particle cross sections above 500 MeV/n and 100 MeV/n respectively are not supported by TOPAS-nBio, values for the higher energy GCRSim fluences were obtained via a curve-fitting procedure. Additional simulations were performed in accordance with the above methodology for protons and $\alpha$-particles at additional energies in order to accumulate enough data points for curve-fitting. For protons, mean values for each scored quantity were fit using the general relationship
\begin{eqnarray}
    f(x) = a(x - c)^{-b} + d
\end{eqnarray}
for incident particle energy $x$ and non-negative fit parameters $a$, $b$, $c$, and $d$. For $\alpha$-particles, which exhibited larger variability in the scored quantities, the simplified power-law function
\begin{eqnarray}
    g(x) = ax^{-b}
\end{eqnarray}
was used to reduce uncertainty in the curve-fitting process. Mean values of the scored quantities were then extrapolated along these curves to match the 1000 MeV/n proton and 150, 250, and 1000 MeV/n $\alpha$-particle fluences in the GCRSim spectrum. Standard deviations for the extrapolated quantities were computed from uncertainties in the fit parameters using standard error propagation techniques. 

The measured values of each of the scored quantities depend on the total fluence simulated. In order to relate the simulated values to their respective values for a given dose of GCRSim, scale factors were calculated for each ion-energy combination. For a dose $z$ of GCRSim, the scale factor $\sigma_{ij}^z$ for ion $i$ and energy $j$ in the fluence distribution was calculated as
\begin{eqnarray}
    \sigma_{ij}^z = D_{ij}^z / D_{ij}^s
\end{eqnarray}
where $D_{ij}^z$ is the fractional dose of GCRSim scaled to total dose $z$ and $D_{ij}^s$ is the corresponding simulated dose to the whole neuron averaged over the pooled runs. As such, the mean simulated value of a scored quantity $q_{ij}^s$ can be related to the scaled value $q_{ij}^z$ as
\begin{eqnarray}
    q_{ij}^z = \sigma_{ij}^z q_{ij}^s
\end{eqnarray}
for each ion-energy combination. Values summed over the entire GCRSim fluence distribution can be calculated as
\begin{eqnarray}
    Q^z = \sum_{ij} q_{ij}^z.
\end{eqnarray}
Standard deviations were scaled appropriately using standard error propagation formulas.

\subsection{Calculation of Dendritic Hit Curves}

In addition to scoring energy deposition and fluence quantities, the cumulative fraction of dendritic segments irradiated was tracked as a function of the whole neuron absorbed dose for each ion and energy. This was achieved by tabulating the unique volume segments traversed during each of the 100 successive runs for each ion-energy combination. These relationships were fit using the functional form
\begin{eqnarray}
    h_{ij}(y) = 1 - \e^{\tau_{ij} y}
\end{eqnarray}
where $y$ is the neuronal absorbed dose and $\tau$ is a fit parameter describing the saturation of the curve. In order to generate hit curves $h(y)$ for the unsupported proton and $\alpha$-particle energies, the fit parameter $\tau$ itself was fit as a function of primary particle energy $x$ according to
\begin{eqnarray}
    \tau_k(x) = m_k x^{n_k}
\end{eqnarray}
for new fit parameters $m_k$ and $n_k$ where $k \in \{\mathrm{p}, \alpha\}$ indexes either protons or $\alpha$-particles. The values of $\tau_{\mathrm{p}}(x)$ at $x = 1000$ MeV/n and $\tau_{\alpha}(x)$ at $x = 150$, 250, and 1000 MeV/n were extrapolated and used to fit the corresponding hit curves. The overall fraction of dendritic segments irradiated as a function of GCRSim dose $z$ was then obtained from standard probability theory. Denoting the joint probability of irradiation by an ion species $i$ of any energy $j$ in the GCRSim spectrum as
\begin{eqnarray}
    H_i(z) = 1 - \prod_j (1 - h_{ij}(D_{ij}^z)),
\end{eqnarray}
the cumulative hit probability across the entire GCRSim fluence spectrum at total dose $z$ was calculated as
\begin{eqnarray}
    H(z) = 1 - \prod_i (1 - H_i (z)).
\end{eqnarray}

\section{Results}

Curve fitting results for protons and $\alpha$-particles are given in supplementary figures S1-S6. Low energy particles ($<1$ MeV/n) are stopped by the surrounding water environment before reaching the neuron. Simulated quantities peak near a primary particle energy of $5-7$ MeV/n and subsequently decay approximately hyperbolically at larger energies. As the energy of the primary particles increases towards the peak value, more secondary particles are produced which increases the frequency of energy deposition events (supplementary figure S6). However, as the energy of the primaries passes the peak value, the energy deposition Bragg peak moves outside the simulation volume and the total energy deposition declines. For the GCRSim and SimGCRSim spectra considered here, all primary particle energies exceed this peak value. As such, the lower energy proton and $\alpha$-particle beams contribute more damage per Gy of absorbed dose.

\subsection{Nano-Dosimetry and Fluence Statistics}

All simulated quantities were scaled to a GCRSim neuronal reference dose of $z = 0.5$ Gy. Although the reference dose is arbitrary, it was chosen to match the dose being used in initial GCRSim experiments and corresponds to approximately the dose an astronaut would experience during a $2-3$ year Mars mission [3]. Figure 2 shows the fractional dose absorbed by the neuron and various subcellular structures for each primary ion species summed over its respective energies. By definition, mean values for the whole neuron reflect the exact proportions of the GCRSim fluence distribution for each ion ($\sum_j D_{ij}^z$; sums of entries for each ion in table 1). The total absorbed doses across all particles to the soma, dendrites, and spines are $0.54 \pm 0.09$ Gy, $0.47 \pm 0.02$ Gy, and $0.8 \pm 0.5$ Gy, respectively. Standard deviations in the soma and dendritic doses reflect the stochastic nature of Monte Carlo transport. The larger standard deviation of the spine dose is due to the additional stochasticity in the volumes of the spines themselves. Variations in the mean subcellular volume doses away from 0.5 Gy reflect inhomogeneities in the irradiation profile. In general, as the fluence dose increases, the energy deposition scored in each volume becomes more homogeneous and the absorbed dose approaches the fluence dose. Smaller volumes require larger fluence doses to achieve homogeneity [32]. As such, the average spine dose is more susceptible to local variations in track structure. For the larger soma and dendrite volumes, these variations are coarse-grained and the compartmental doses are similar to the neuronal dose.

\begin{figure}
    \includegraphics[width=\textwidth]{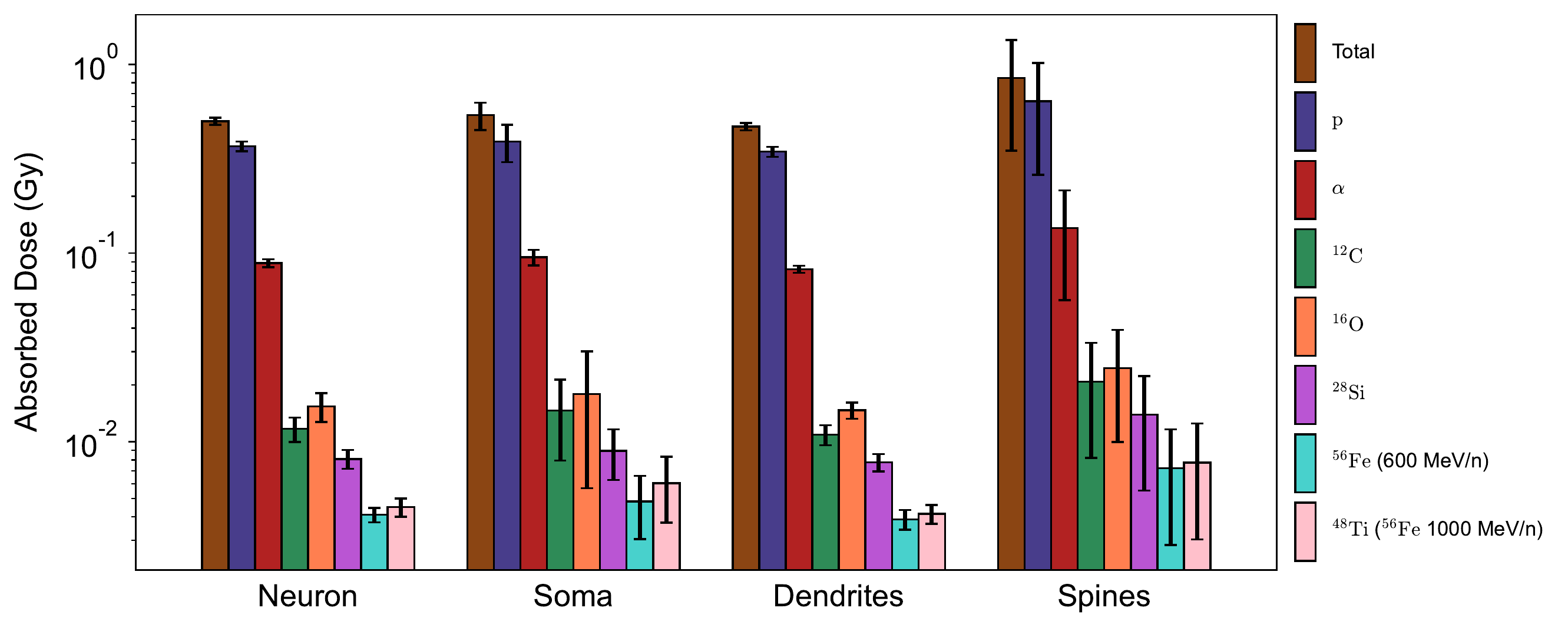}
    \caption{Doses to the neuron and subcellular structures after GCRSim irradiation. Values are scaled to a total neuronal absorbed dose of 0.5 Gy GCRSim. Colors indicate the fractional contributions from different ion beams in the GCRSim fluence distribution. The $^{48}$Ti fluence was replaced with 1000 MeV/n $^{56}$Fe because $^{48}$Ti cross sections are not defined in Geant4-DNA. From left to right, the ion contributions are: total GCRSim (brown), proton (p; navy), $\alpha$-particle ($\alpha$; red), $^{12}$C (green), $^{16}$O (orange), $^{28}$Si (purple), $^{56}$Fe (600 MeV/n; turquoise), $^{56}$Fe (1000 MeV/n, replacing $^{48}$Ti; pink). Error bars indicate mean $\pm$ standard deviation as discussed in the text.}
    \label{fig:2}
\end{figure}

A further analysis of the energy deposition by spine type is presented in figure 3. Mushroom spines, despite accounting for only $31 \pm 7\%$ of all spines, receive $78 \pm 3\%$ of the total spine energy deposition ($1.0 \pm 0.1$ MeV) due to their larger average volume. Thin and stubby spines receive $17 \pm 3\%$ ($0.22 \pm 0.05$ MeV) and $5 \pm 2\%$ ($0.07 \pm 0.02$ MeV) of the total spine energy deposition, respectively. 

\begin{figure}
    \includegraphics[width=\textwidth]{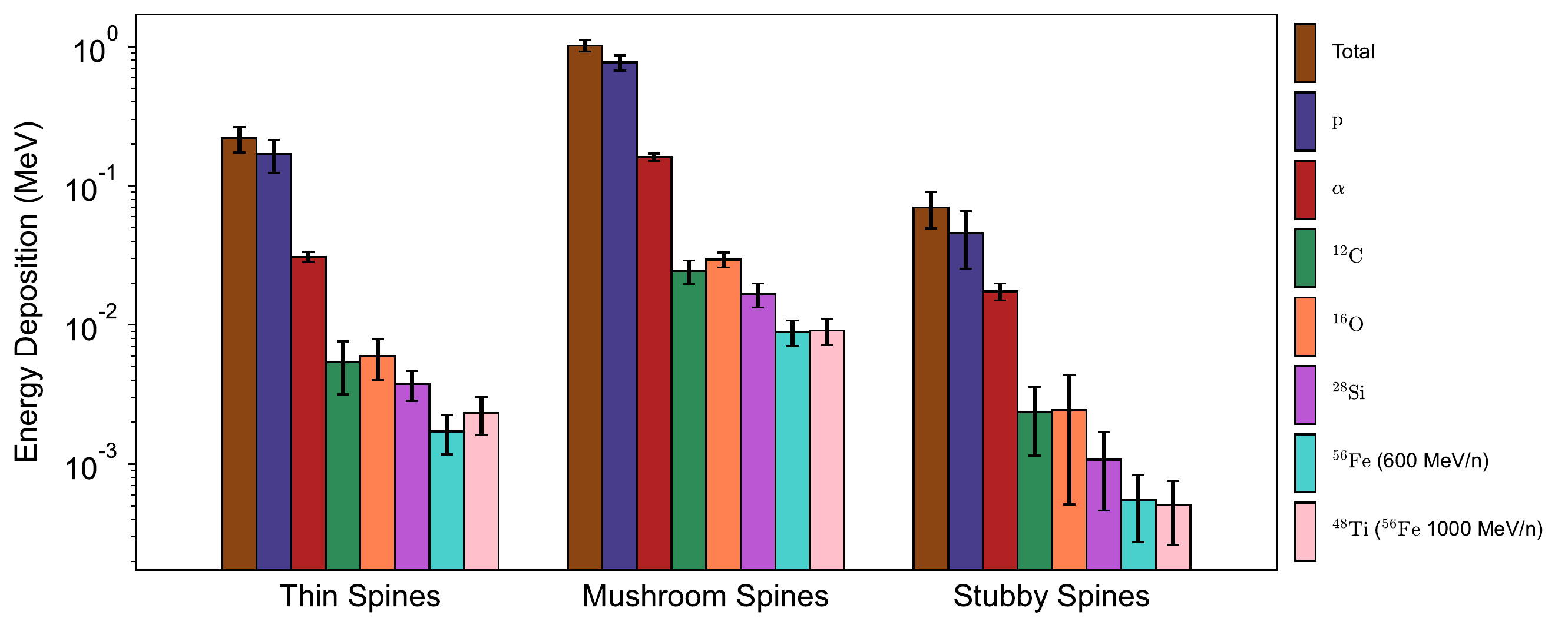}
    \caption{Energy deposition in thin, mushroom, and stubby spines after irradiation with GCRSim. Values are scaled to a total neuronal absorbed dose of 0.5 Gy GCRSim. Colors indicate the fractional contributions from different ion beams in the GCRSim fluence distribution. The $^{48}$Ti fluence was replaced with 1000 MeV/n $^{56}$Fe because $^{48}$Ti cross sections are not defined in Geant4-DNA. From left to right, the ion contributions are: total GCRSim (brown), proton (p; navy), $\alpha$-particle ($\alpha$; red), $^{12}$C (green), $^{16}$O (orange), $^{28}$Si (purple), $^{56}$Fe (600 MeV/n; turquoise), $^{56}$Fe (1000 MeV/n, replacing $^{48}$Ti; pink). Error bars indicate mean $\pm$ standard deviation as discussed in the text.}
    \label{fig:3}
\end{figure}

Figure 4 depicts the relative contributions of different physics processes to the energy deposition in the full neuron volume. The dominant contribution is from ionizations, which account for $68 \pm 1\%$ of the total deposited energy. This is followed by smaller contributions from electron solvations ($21 \pm 1\%$), electronic excitations ($6.7 \pm 0.5\%$), vibrational excitations ($3.6 \pm 0.2\%$), and electron attachments ($0.72 \pm 0.05\%$). It is interesting to compare these values with the frequencies at which such processes occur (figure 5). In particular, vibrational excitations, despite their minimal contribution to the overall energy deposition, occur at nearly five times the rate as ionizations and account for $70 \pm 2\%$ of all energy deposition events. The next most frequent processes are ionizations ($14 \pm 1\%$), electron solvations ($14 \pm 1\%$), electronic excitations ($1.8 \pm 0.2\%$), and electron attachments ($0.23 \pm 0.02\%$).

\begin{figure}
    \includegraphics[width=\textwidth]{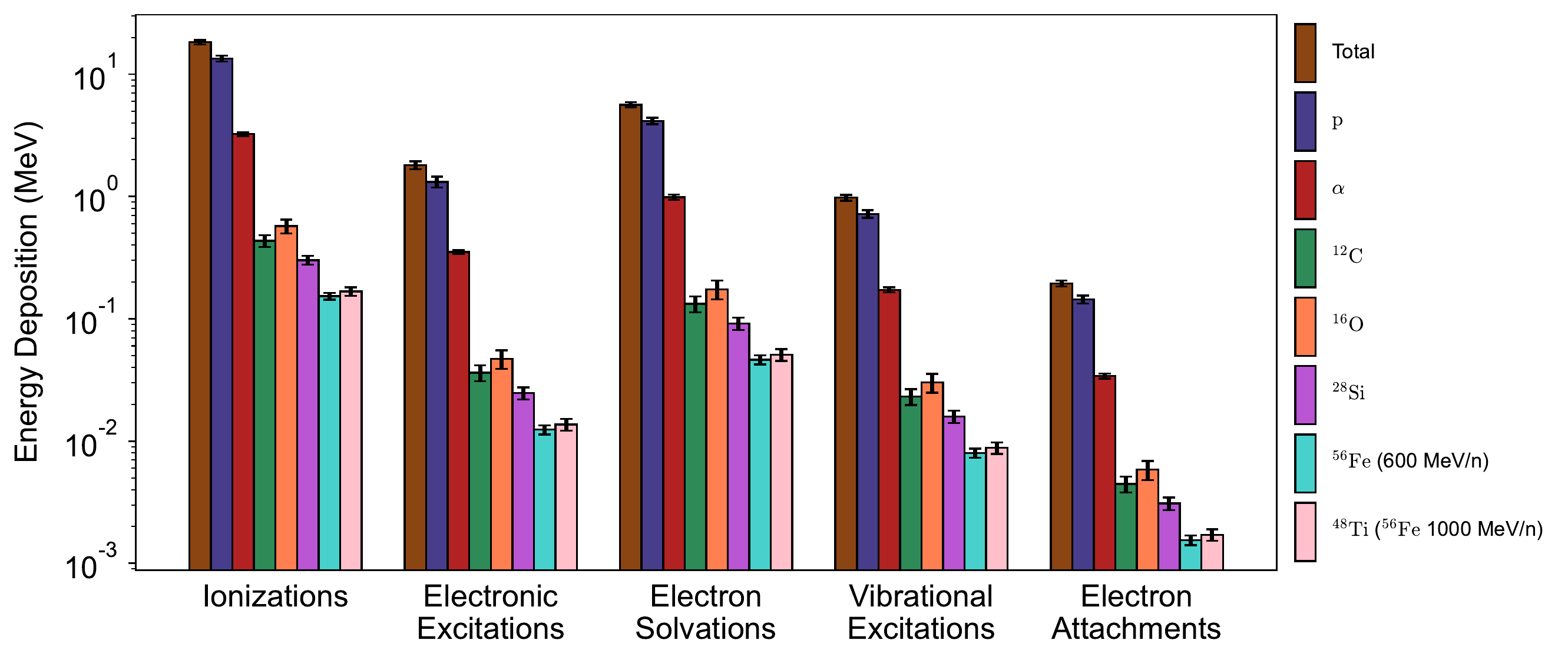}
    \caption{Contributions of different physics processes to neuronal energy deposition during irradiation with GCRSim. Values are scaled to a total neuronal absorbed dose of 0.5 Gy GCRSim. Colors indicate the fractional contributions from different ion beams in the GCRSim fluence distribution. The $^{48}$Ti fluence was replaced with 1000 MeV/n $^{56}$Fe because $^{48}$Ti cross sections are not defined in Geant4-DNA. From left to right, the ion contributions are: total GCRSim (brown), proton (p; navy), $\alpha$-particle ($\alpha$; red), $^{12}$C (green), $^{16}$O (orange), $^{28}$Si (purple), $^{56}$Fe (600 MeV/n; turquoise), $^{56}$Fe (1000 MeV/n, replacing $^{48}$Ti; pink). Error bars indicate mean $\pm$ standard deviation as discussed in the text.}
    \label{fig:4}
\end{figure}

\begin{figure}
    \includegraphics[width=\textwidth]{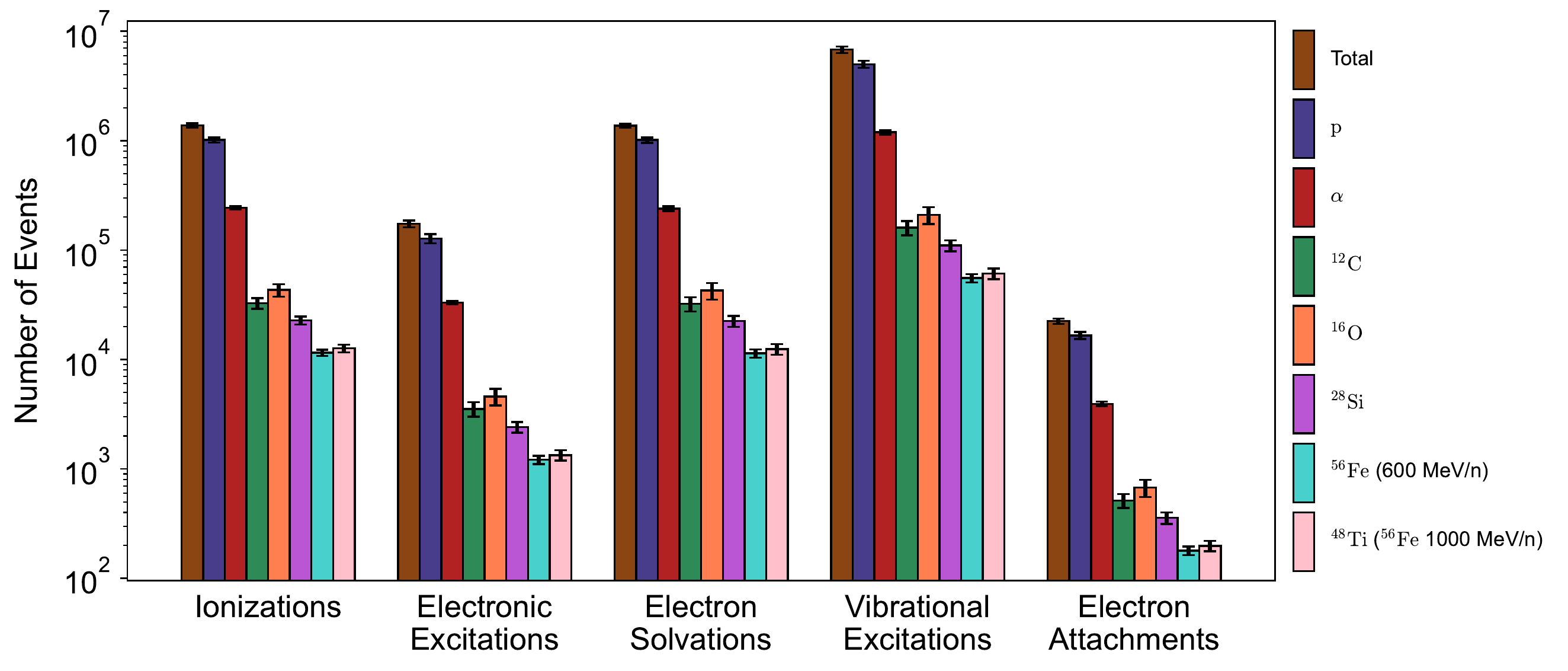}
    \caption{Frequencies of different physics processes during irradiation with GCRSim. Values are scaled to a total neuronal absorbed dose of 0.5 Gy GCRSim. Colors indicate the fractional contributions from different ion beams in the GCRSim fluence distribution. The $^{48}$Ti fluence was replaced with 1000 MeV/n $^{56}$Fe because $^{48}$Ti cross sections are not defined in Geant4-DNA. From left to right, the ion contributions are: total GCRSim (brown), proton (p; navy), $\alpha$-particle ($\alpha$; red), $^{12}$C (green), $^{16}$O (orange), $^{28}$Si (purple), $^{56}$Fe (600 MeV/n; turquoise), $^{56}$Fe (1000 MeV/n, replacing $^{48}$Ti; pink). Error bars indicate mean $\pm$ standard deviation as discussed in the text.}
    \label{fig:5}
\end{figure}

In response to 0.5 Gy of GCRSim, we find an average of $1760 \pm 90$ energy deposition events per micron of dendritic length, $250 \pm 10$ of which are ionizations. We also report an additional $130 \pm 10$ events per dendritic spine, including $19 \pm 2$ ionizations per spine. For mushroom, thin, and stubby spines separately, we find an average of $330 \pm 80$, $50 \pm 20$, and $30 \pm 10$ energy deposition events per spine, respectively. For ionizations, the values are $50 \pm 10$, $7 \pm 2$, and $4 \pm 2$ per spine.    

Furthermore, we report the fractional energy deposition in the neuron as a function of mediating particle species (figure 6). As expected, each primary ion type deposits the majority of its energy into the neuron via secondary electrons, which account for $77 \pm 1\%$ of the total energy deposition. Smaller contributions come from $^1$H ($17 \pm 1\%$) and $^4$He ($4.1 \pm 0.3\%$), as well as $^{12}$C, $^{16}$O, $^{28}$Si, $^{56}$Fe, and $\gamma$-radiation ($< 1\%$ each). 

\begin{figure}
    \includegraphics[width=\textwidth]{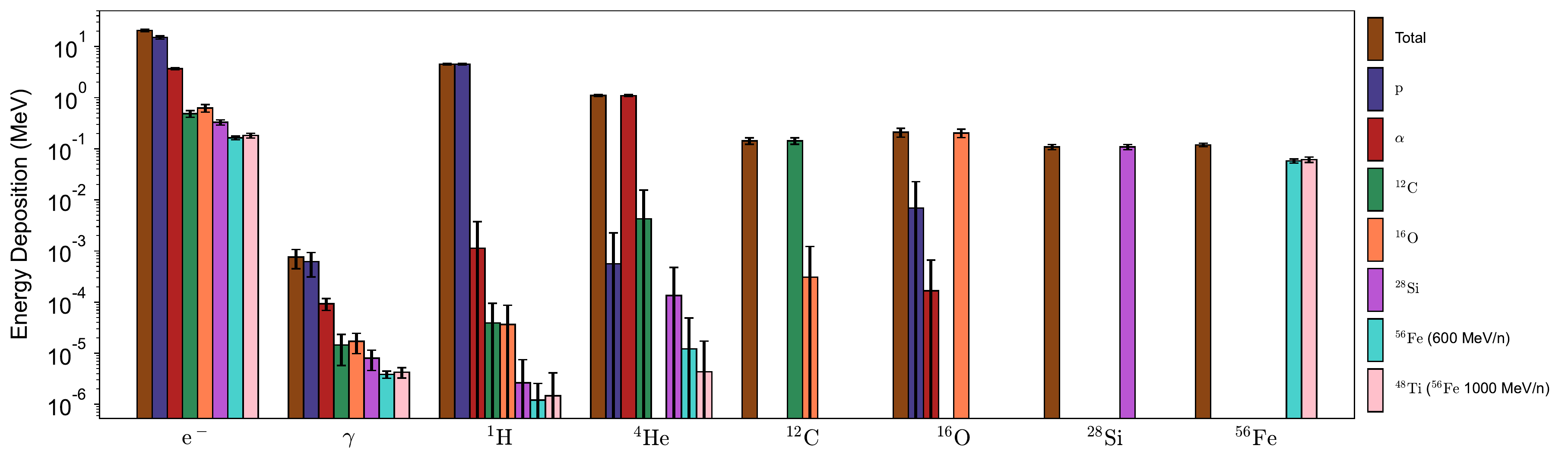}
    \caption{Contributions of different particle species to neuronal energy deposition during irradiation with GCRSim. Values are scaled to a total neuronal absorbed dose of 0.5 Gy GCRSim. The x-axis indicates the particle species mediating the energy deposition while different colors refer to the primary particle initiating the track (different ion beams in the GCRSim fluence distribution). $^1$H includes protons and minor contributions from neutral hydrogen atoms. $^4$He includes $\alpha$-particles and contributions from neutral and singly ionized helium atoms. The $^{48}$Ti fluence was replaced with 1000 MeV/n $^{56}$Fe because $^{48}$Ti cross sections are not defined in Geant4-DNA. From left to right, the ion contributions are: total GCRSim (brown), proton (p; navy), $\alpha$-particle ($\alpha$; red), $^{12}$C (green), $^{16}$O (orange), $^{28}$Si (purple), $^{56}$Fe (600 MeV/n; turquoise), $^{56}$Fe (1000 MeV/n, replacing $^{48}$Ti; pink). Error bars indicate mean $\pm$ standard deviation as discussed in the text.}
    \label{fig:6}
\end{figure}

\subsection{Dendritic Hit Curves}

We also assessed the spatial distribution of energy deposition events throughout the dendrites. The fitted hit curves $h_{ij}(y)$ of equation 6 are plotted in figure 7. These functions quantify the cumulative fraction of dendritic segments hit by at least one particle (or equivalently, the probability of hitting a given segment) as a function of neuronal absorbed dose for each of the 33 GCRSim ion-energy combinations. Figure 7 demonstrates that, for protons and $\alpha$-particles, the likelihood of irradiating a particular dendritic segment increases with increasing particle energy. This trend is particularly noticeable at low doses, where the hit curves are far from saturation and approximately linear. For example, irradiation of 0.01 Gy by 1000 MeV protons registers an 85\% dendritic segment hit probability, compared to just 44\% for the same dose of 250 MeV protons. This difference diminishes towards zero with increasing dose as both hit curves approach unity. The quantitative features of these hit curves are captured succinctly by the fit parameters $m_k$ and $n_k$ of equation 7 which describe how the characteristic dose, $1/ \tau(x)$, changes as a function of incident particle energy. Here, $1/ \tau$ is defined as the absorbed dose at which 63\% of segments are irradiated. For protons and $\alpha$-particles, $\tau(x)$ is described well by a power-law relationship (supplementary figure S7). Values of $\tau_{ij}$ for each of the 33 GCRSim beams are given in supplementary table S2. 

\begin{figure}
    \includegraphics[width=\textwidth]{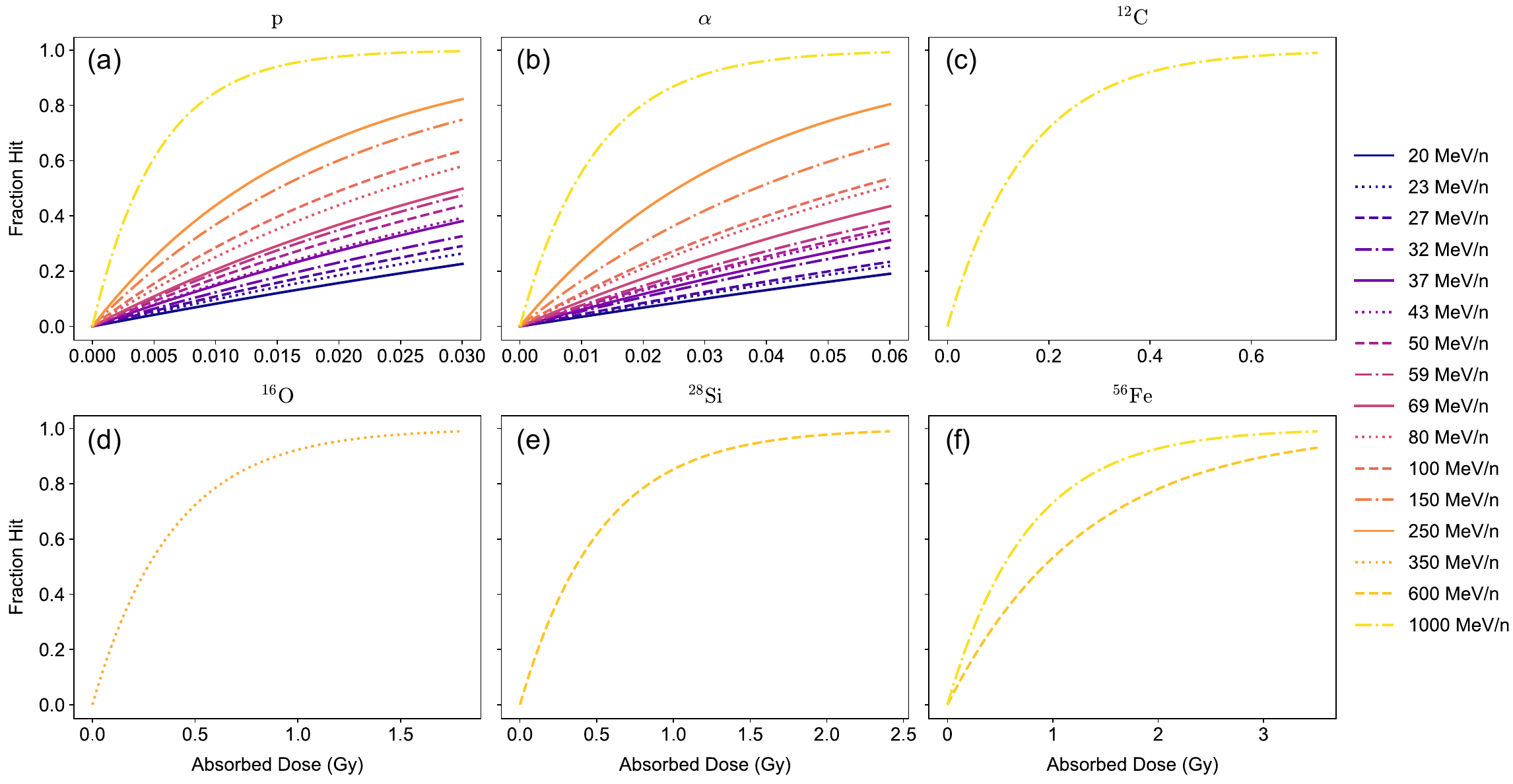}
    \caption{Dendritic hit curves for each GCRSim ion-energy beam. The x-axis of each subplot denotes neuronal absorbed dose and the y-axis plots the fraction of dendritic segments receiving at least one energy deposition event during irradiation. Ion species are indicated by subplot headers: \textbf{(a)} proton, \textbf{(b)} $\alpha$-particle, \textbf{(c)} $^{12}$C, \textbf{(d)} $^{16}$O, \textbf{(e)} $^{28}$Si, \textbf{(f)} $^{56}$Fe. Energies are indicated by different colors (line styles varied for improved visibility).}
    \label{fig:7}
\end{figure}

The curves illustrated in figure 7 demonstrate a reduction in hit probability with increasing LET and are consistent with previous findings [30]. The observed trends can be understood by considering two competing physical mechanisms. The hit probability depends on both the fluence of the incident particles, as well as their penetration depth. At lower energies, a greater number of primary particles is required to deliver a given absorbed dose. For an isotropic fluence, increasing the number of primary particles facilitates a more homogeneous irradiation profile, thus increasing access to new regions of the neuron and raising the hit fraction. Furthermore, the total number of energy deposition events declines as $\sim x^{-0.81}$ for protons and  $\sim x^{-0.98}$ for $\alpha$-particles with increasing energy $x$ for all GCRSim energies (supplementary figure S6). However, higher energy primary particles produce higher energy secondaries, which are able to traverse greater distances before coming to rest inside the simulation volume. It is this latter effect which drives the proportional relationship between hit probability and incident energy. By contrast, an inverse relationship is seen when comparing hit probabilities between different particles with increasing charge. Higher charge particles make more densely clustered tracks, depositing energy within a more tightly-spaced volume. As such, the neuron accumulates a much larger dose than it would with a lower charge particle for a given hit fraction. 

The cumulative hit curve $H(z)$ for the full GCRSim spectrum is displayed in figure 8. The steepness of the curve is dictated almost entirely by protons, as a consequence of their low LET and dominant contribution to the total fluence dose. The exact functional form of the GCRSim hit curve is given by equations 8 and 9. Noting the similarities between this curve and those shown in figure 7, an effective saturation parameter can be deduced by fitting the simpler function given by equation 6. The characteristic dose is found to be $1/ \tau_{\mathrm{GCR}} = 0.014$ Gy, similar to that of the highest energy protons and $\alpha$-particles. Also shown in figure 8 are the primary particle hit curves $H_i(z)$ denoting the fraction of segments irradiated by a given primary particle track. We find that exposure to GCRSim results in irradiation by protons of nearly all dendritic segments at very low doses (effective $1/ \tau$ value of 0.016 Gy). Widespread irradiation by $\alpha$-particles is also likely at spaceflight-relevant doses (effective $1/ \tau$ of 0.17 Gy). By contrast, irradiation by higher LET ions remains sparse even at much larger exposures and declines in likelihood with increasing LET. The effective $1/ \tau$ values for $^{12}$C, $^{16}$O, $^{28}$Si, $^{48}$Ti (substituted with 100 MeV/n $^{56}$Fe), and 600 MeV/n $^{56}$Fe tracks are 6.8, 13, 32, 85, and 160 Gy, respectively.

\begin{figure}
    \includegraphics[width=\textwidth]{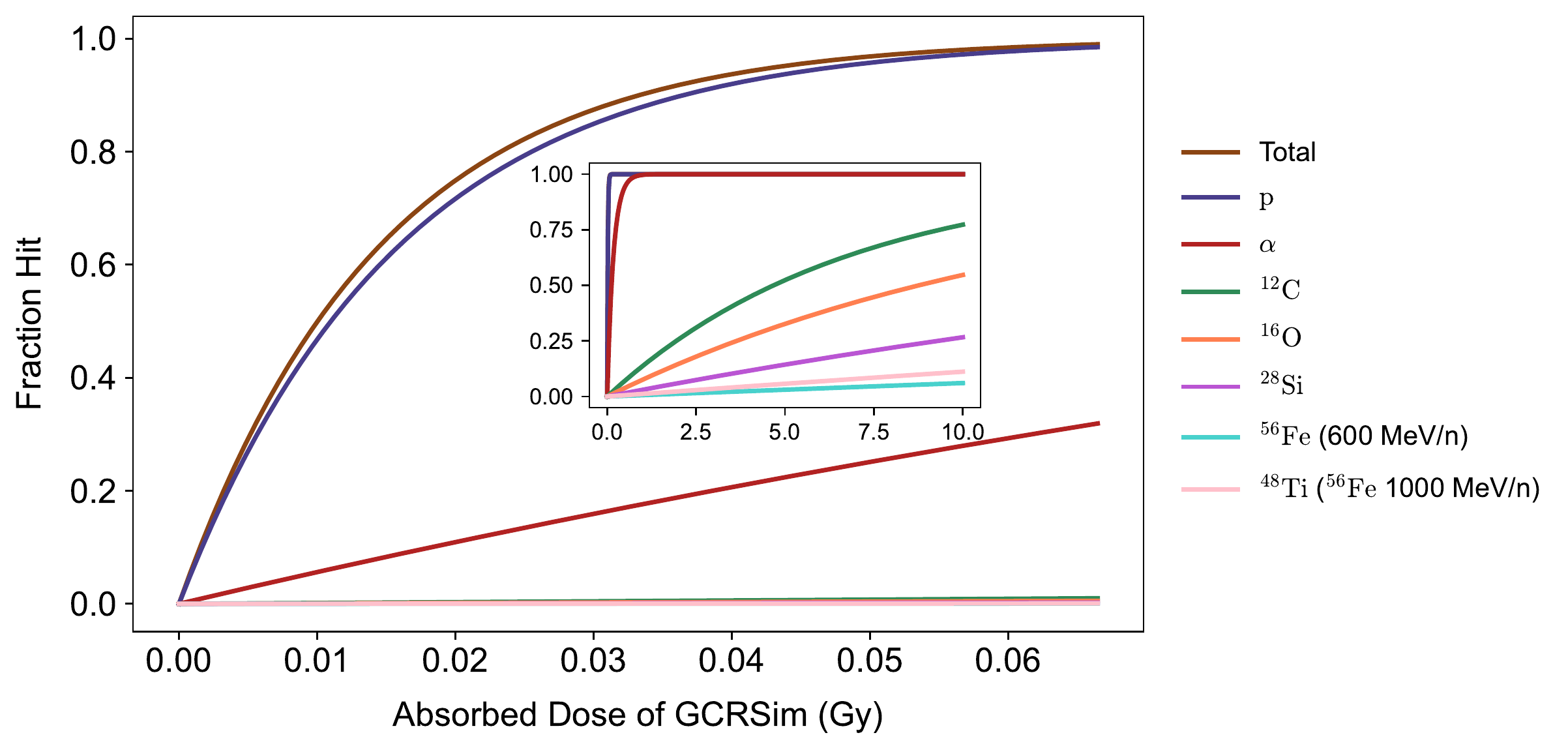}
    \caption{Cumulative and primary particle dendritic hit curves as a function of total neuronal absorbed dose of GCRSim. Colors indicate the fractional contributions from different ion beams in the GCRSim fluence distribution. The $^{48}$Ti fluence was replaced with 1000 MeV/n $^{56}$Fe because $^{48}$Ti cross sections are not defined in Geant4-DNA. The ion contributions are: total GCRSim (brown), proton (p; navy), $\alpha$-particle ($\alpha$; red), $^{12}$C (green), $^{16}$O (orange), $^{28}$Si (purple), $^{56}$Fe (600 MeV/n; turquoise), $^{56}$Fe (1000 MeV/n, replacing $^{48}$Ti; pink). \textbf{(Inset)} The same graph plotted on a wider dose range.}
    \label{fig:8}
\end{figure}

\subsection{GCRSim versus SimGCRSim}

A major challenge of radiobiology studies is the interpretation of diverse experimental results which utilize different ion-energy combinations. Early studies using single beam irradiations revealed a multitude of complex biological responses which vary by dose, particle species, energy, and LET [7]. Although the reasons for these differences are not completely understood, differences in track structure—and in particular, ionization density—are thought to be a dominant factor. 

In order to facilitate comparisons between future experiments and to explore the differences between realistic and simplified fluence distributions, we compare the results of the previous two sections to those obtained using the simplified spectrum of SimGCRSim. Overall, we find remarkably similar results between the 33 beam GCRSim and the 6 beam SimGCRSim. Supplementary table S3 summarizes the relative contributions of different physics processes to the neuronal energy deposition for both spectra. There is no statistically significant difference in the frequency of any process between the two fluence distributions. Similarly, there is no significant difference in the density of energy deposition or ionization events in any of the subcellular volumes, including the dendrites and each of the three spine types (supplementary table S4). 

Figure 9 shows the cumulative and primary particle dendritic hit curves for the SimGCRSim spectrum. The effective characteristic dose for the full spectrum is $1/ \tau_{\mathrm{SIM}} = 0.011$ Gy, similar to that for GCRSim. The respective values for proton, $\alpha$-particles, $^{16}$O, $^{28}$Si, and $^{56}$Fe tracks are 0.011, 0.20, 6.5, 52, and 130 Gy, respectively. 

\begin{figure}
    \includegraphics[width=\textwidth]{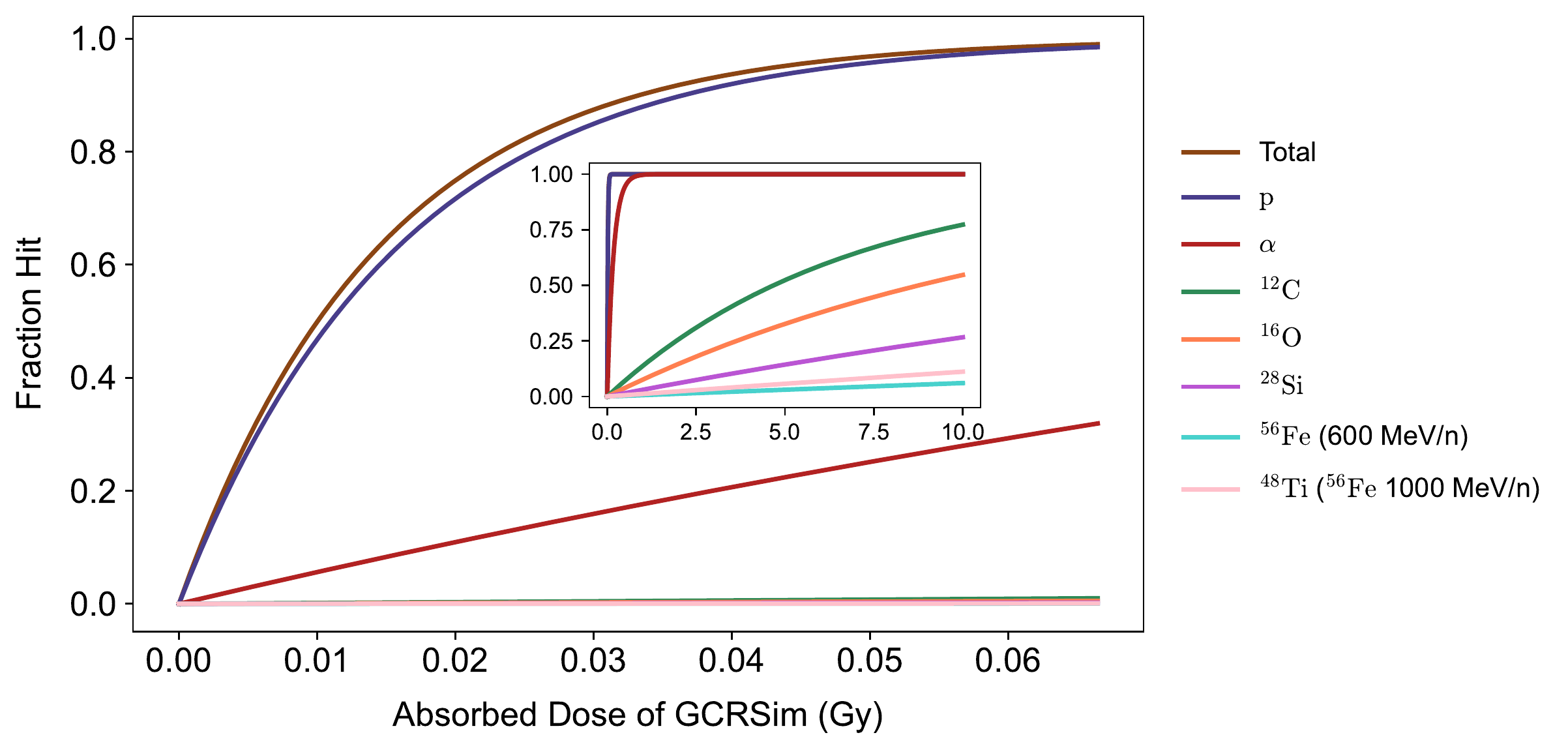}
    \caption{Cumulative and primary particle dendritic hit curves as a function of total neuronal absorbed dose of SimGCRSim. Colors indicate the fractional contributions from different ion beams in the SimGCRSim fluence distribution. The ion contributions are: total SimGCRSim (brown), proton (p; navy), $\alpha$-particle ($\alpha$; red), $^{16}$O (orange), $^{28}$Si (purple), $^{56}$Fe (turquoise). \textbf{(Inset)} The same graph plotted on a wider dose range.}
    \label{fig:9}
\end{figure}

\section{Discussion}

Our dosimetry results suggest that mushroom spines are at a greater risk for radiation-induced damage relative to other spine types. Because of their larger average volume, mushroom spines are susceptible to more frequent ionizations and larger energy depositions. Indeed, changes in mushroom spine populations have been reported experimentally in adult mice at multiple timepoints post-irradiation with particles of varying LET. One study using 600 MeV/n $^{16}$O demonstrated reductions in mushroom spine density on basal CA1 dendrites and dentate gyrus cells with doses of 0.1, 0.25, and 1 Gy at two weeks post-irradiation [12]. In multiple treatment groups, such changes were present in the absence of reductions in other spine densities [12]. Similar results have been shown with 10 Gy $\gamma$-radiation at one week post-irradiation in CA1 basal dendrites and at both one week and one month post-irradiation in dentate gyrus cells [26]. In response to 1 Gy of 150 MeV protons, reductions in mushroom spine densities have been noted in both apical and basal CA1 dendrites, apical CA3 dendrites, and dentate gyrus cells as late as nine months post-irradiation [19]. Increases in mushroom spine proportions have also been reported in apical CA1 dendrites one week and one month post-irradiation with 10 Gy $\gamma$-radiation [26] and in dentate gyrus cells three months post-irradiation with 0.5 Gy of 600 MeV/n $^{56}$Fe [20]. Occasionally, increases in the density of one spine type have been reported alongside simultaneous decreases in another, potentially indicating a type of compensatory mechanism within the cell [12], [19], [20]. Other studies, however, have failed to find changes in mature spine populations altogether. Parihar et al. reported no significant alterations to dentate gyrus cell mushroom, thin, or stubby spines at both 10 days and 30 days post-irradiation with either 250 MeV protons (0.1 or 1 Gy) [15] or $\gamma$-radiation (1 or 10 Gy) [49]. The diversity of these results reflects the inherent heterogeneity of experimental protocols, which vary in dose, particle species, energy, timepoint, and brain region. A further consideration for future studies is dose rate. Mushroom spines typically persist for months and therefore accumulate significantly larger doses during chronic exposures as compared to thin spines, which exist transiently only for several days [20]. Long-duration studies with GCRSim will be helpful to standardize these variables.

The density of energy deposition events along each dendrite may be an important factor in determining neuronal outcome. It is possible that while sparsely spaced lesions are repaired without altering cellular function, more dense damage clusters are able to overwhelm the cell and lead to reductions in dendritic arborization and spine pruning. Ionizations are particularly harmful as they inflict the largest energy deposition per event. Here we have reported the average number of energy deposition events and ionizations per micron of dendritic length and per dendritic spine for each of the three spine types. Such results could be incorporated into existing models of radiation-induced morphological changes. Recently, Alp and Cucinotta developed a theoretical model of dendritic snipping as a function of incident particle LET [50]. The fluence data presented here could be used to extend that model to include the influence of track structure and different physics processes in a probabilistic way. Further studies could incorporate these relationships into more complex models which include damage repair kinetics [51]. On the experimental side, these data could be combined with future \textit{in vitro} studies of dendritic remodeling to deduce possible correlations between event density and structural changes in specific cells. 

The observed trend of increasing dendritic segment hit probability with increasing particle charge is consistent with other computational studies comparing neuronal hit curves with proton, $^{12}$C, and $^{56}$Fe beams [29]–[32]. Similar inverted exponential relationships between hit percentage and neuronal absorbed dose have been observed [29]. The shapes of these curves demonstrate that at low doses, a small increase in absorbed dose corresponds to a significantly larger increase in the hit probability than an equivalent increase in the high-dose saturated regime. Whereas the low-dose regions of the curves follow an approximately linear trend, the high-dose regions asymptote as the probability approaches unity. Because hit probability is inversely proportional to compartment volume, smaller structures exhibit linear regimes that extend to higher doses. For the $5.02 \pm 2.76$ $\mu$m$^3$ dendritic segments used in our study, hit curve saturation was realized for relatively low doses of GCRSim, reaching a value of 99\% at 0.066 Gy. Much larger doses are likely necessary for complete irradiation of smaller structures such as spines ($\mu$m scale), synaptic receptors (100 nm scale), and ion channels (10 nm scale). Indeed, a study by Bayarchimeg et al. demonstrated that hit curves for protons, $^{12}$C, and $^{56}$Fe beams scored in dendritic spines only approach unity for doses larger than 1 Gy, while those for synaptic receptors remain linear up to at least 4 Gy [29]. 

The observed functional behavior of subcellular hit curves may have important consequences for estimating the risk of radiation-induced cognitive impairment. Receptors and ion channels control the functional behavior of neurons by facilitating synaptic transmission and regulating membrane potential. Linearity of the receptor hit curves up to high doses suggests that extended deep-space missions may have a cumulative impact on neuronal health. As dose increases, so too does the number of irradiated receptors in a proportional manner. Nevertheless, it is still unclear which aspect of irradiation (fluence, dose, LET, etc.) is the dominant factor mediating neuronal outcome. Future studies on the relationship between receptor hits and receptor deactivation will be critical for translating these fluence statistics into risk assessment models. Such studies would help fill in the missing link between Monte Carlo modeling of physical parameters and radiobiological studies of cellular dysfunction. Relationships between fluence quantities and receptor deactivation could then be integrated with patch-clamp measurements into biophysical Hodgkin-Huxley-type models to directly estimate changes in electrophysiological parameters as a function of dose. As such, experimental studies on the low-dose damage responses of individual receptors and synapses are essential for formulating a mechanistic model of radiation-induced changes to neuronal function. 

Our comparisons between GCRSim and SimGCRSim suggest that the average physical responses to equal doses of each fluence distribution are nearly identical. This is found to be true for both the frequency and total energy deposition of each physics process. Nevertheless, further examination of the spatial distributions of energy deposition events at the scale of individual spines is needed. Additionally, we observed only minor differences between the dendritic segment hit probabilities calculated with GCRSim ($1/ \tau_{\mathrm{GCR}} = 0.014$ Gy) and SimGCRSim ($1/ \tau_{\mathrm{SIM}} = 0.011$ Gy). Whether these differences are significant across multiple statistical runs and neuron geometries could be explored in a future work. 

To date, there have been very few published studies exploring the effects of multi-ion irradiation on CNS function. We are aware of only one experimental study which examined the electrophysiological and behavioral changes induced by the 6 beam SimGCRSim spectrum and none using the full 33 beam GCRSim. The authors of the study demonstrated enhanced inhibitory synaptic signaling within the CA1 region and corresponding impairments to learning and memory behavior $1-2$ months post-irradiation with 0.3 Gy of SimGCRSim [52]. Only minor electrophysiological differences were observed when comparing the results to a similar 5 beam variation of the fluence distribution [52]. However, whether there are significant differences in the biological responses to different mixed-ion radiation fields is still debated. Another study using a slightly different 6 beam spectrum reported notable differences in locomotor activity, open field measures of anxiety, forced swim tests, and passive avoidance memory tasks when compared to a previous study using a 3 beam radiation field at identical doses [53], [54]. Examination of additional time points and responses to more complex fluence distributions are warranted.

An important aspect which was not addressed in this work is the influence of radiolysis products during the chemical stage of track development. Between $10^{-12}$ and a few $10^{-9}$ seconds post-irradiation, ionized and dissociated water molecules react to form reactive oxygen species. These products diffuse throughout the cellular environment, leading to additional energy deposition and ionization events until they are scavenged by other molecules. Single ion simulations of the chemical stage have shown that both the total number of produced chemical species and their flux incident upon realistic ion channel geometries increase with particle LET up to 1 MeV $\mu$m$^{-1}$ [29], [30]. The presence of these radiolytic products increases the effective damage radius of each particle track and may enhance the total damage inflicted upon the neuron [30]. However, no studies to date have performed such simulations using the full GCRSim or SimGCRSim spectra. Moreover, the event thresholds which are necessary to initiate damage to individual neuronal structures are unknown. Further studies specifically assessing these nano-scale parameters with realistic fluence distributions may be useful for interpreting or designing future experiments examining oxidative stress or anti-oxidative biological countermeasures. 

\section{Conclusion}

The present work provides an in-depth assessment of the nano-scale physics associated with neuronal irradiation by the GCRSim and SimGCRSim beams in use at the NASA Space Radiation Laboratory. To our knowledge, this is the first study to quantify and compare such parameters in a realistic neuron geometry using these fluence distributions. We have presented a complete documentation of energy deposition events parsed by volume, physics process, and mediating particle species for each ion-energy combination. We have also produced a set of analytical relationships relating a variety of dosimetric and fluence quantities to incident particle energy and absorbed dose. Lastly, we have demonstrated that the 33 beam GCRSim and the 6 beam SimGCRSim irradiation profiles produce very similar physical fluences and energy deposition patterns.

\ack

This work was supported in part by NIH/NCI R01 CA187003 (TOPAS-nBio). J. S. Peter was supported in part by Molecular Biophysics Training Grant NIH/NIGMS T32 GM008313.

\section*{References}

\noindent[1] F. A. Cucinotta and E. Cacao, “Risks of cognitive detriments after low dose heavy ion and proton exposures,” Int. J. Radiat. Biol., vol. 95, no. 7, pp. 985–998, Jul. 2019.\\

\noindent[2] M. Durante and F. A. Cucinotta, “Physical basis of radiation protection in space travel,” Rev. Mod. Phys., vol. 83, no. 4, pp. 1245–1281, 2011.\\

\noindent[3]	L. C. Simonsen, T. C. Slaba, P. Guida, and A. Rusek, NASA’s first ground-based galactic cosmic ray simulator: Enabling a new era in space radiobiology research, vol. 18, no. 5. 2020.\\

\noindent[4]	L. W. Townsend, E. N. Zapp, D. L. Stephens, and J. L. Hoff, “Carrington flare of 1859 as a prototypical worst-case solar energetic particle event,” IEEE Trans. Nucl. Sci., vol. 50, no. 6, pp. 2307–2309, 2003.\\

\noindent[5]	J. H. King, “Solar Proton Fluences for 1977-1983 Space Missions,” J. Spacecr. Rockets, vol. 11, no. 6, pp. 401–408, 1974.\\

\noindent[6]	G. A. Nelson, “Space Radiation and Human Exposures, A Primer,” Radiat. Res., vol. 185, no. 4, pp. 349–358, 2016.\\

\noindent[7]	E. Cekanaviciute, S. Rosi, and S. V Costes, “Central Nervous System Responses to Simulated Galactic Cosmic Rays,” Int. J. Mol. Sci., vol. 19, no. 11, p. 3669, 2018.\\

\noindent[8]	R. A. Britten et al., “Exposure to mission relevant doses of 1 GeV/Nucleon 56Fe particles leads to impairment of attentional set-shifting performance in socially mature rats,” Radiat. Res., vol. 182, no. 3, pp. 292–298, 2014.\\

\noindent[9]	V. K. Parihar et al., “Persistent nature of alterations in cognition and neuronal circuit excitability after exposure to simulated cosmic radiation in mice,” Exp. Neurol., vol. 305, pp. 44–55, 2018.\\

\noindent[10]	M. M. Acharya et al., “New concerns for neurocognitive function during deep space exposures to chronic, low dose-rate, neutron radiation,” eNeuro, vol. 6, no. 4, pp. 1–15, 2019.\\

\noindent[11]	O. Miry et al., “Life-long brain compensatory responses to galactic cosmic radiation exposure,” Sci. Rep., vol. 11, no. 1, pp. 4214–4292, 2021.\\

\noindent[12]	H. Carr et al., “Early effects of 16O radiation on neuronal morphology and cognition in a murine model,” Life Sci. Sp. Res., vol. 17, pp. 63–73, 2018.\\

\noindent[13]	V. K. Parihar et al., “Cosmic radiation exposure and persistent cognitive dysfunction,” Sci. Rep., vol. 6, no. 1, p. 34774, 2016.\\

\noindent[14]	J. A. Bellone, E. Rudobeck, R. E. Hartman, A. Szücs, and R. Vlkolinský, “A single low dose of proton radiation induces long-term behavioral and electrophysiological changes in mice,” Radiat. Res., vol. 184, no. 2, pp. 193–202, 2015.\\

\noindent[15]	V. K. Parihar, J. Pasha, K. K. Tran, B. M. Craver, M. M. Acharya, and C. L. Limoli, “Persistent changes in neuronal structure and synaptic plasticity caused by proton irradiation,” Brain Struct. Funct., vol. 220, no. 2, pp. 1161–1171, 2016.\\

\noindent[16]	J. D. Cherry et al., “Galactic cosmic radiation leads to cognitive impairment and increased a$\beta$ plaque accumulation in a mouse model of Alzheimer’s disease,” PLoS One, vol. 7, no. 12, p. e53275, 2012.\\

\noindent[17]	E. Cacao and F. A. Cucinotta, “Meta-analysis of Cognitive Performance by Novel Object Recognition after Proton and Heavy Ion Exposures,” Radiat. Res., vol. 192, no. 5, pp. 463–472, Aug. 2019.\\

\noindent[18]	C. W. Whoolery et al., “Whole-Body Exposure to 28Si-Radiation Dose-Dependently Disrupts Dentate Gyrus Neurogenesis and Proliferation in the Short Term and New Neuron Survival and Contextual Fear Conditioning in the Long Term,” Radiat. Res., vol. 188, no. 5, pp. 612–631, Sep. 2017.\\

\noindent[19]	F. Kiffer et al., “Late effects of 1 H irradiation on hippocampal physiology,” Life Sci. Sp. Res., vol. 17, pp. 51–62, 2018.\\

\noindent[20]	A. R. Allen, J. Raber, A. Chakraborti, S. Sharma, and J. R. Fike, “56Fe Irradiation Alters Spine Density and Dendritic Complexity in the Mouse Hippocampus,” Radiat. Res., vol. 184, no. 6, pp. 586–594, 2015.\\

\noindent[21]	T. T. Huang, Y. Zou, and R. Corniola, “Oxidative stress and adult neurogenesis-Effects of radiation and superoxide dismutase deficiency,” Semin. Cell Dev. Biol., vol. 23, no. 7, pp. 738–744, 2012.\\

\noindent[22]	E. M. Gore et al., “Phase III comparison of prophylactic cranial irradiation versus observation in patients with locally advanced non-small-cell lung cancer: Primary analysis of Radiation Therapy Oncology Group study RTOG 0214,” J. Clin. Oncol., vol. 29, no. 3, pp. 272–278, 2011.\\

\noindent[23]	A. Sun et al., “Phase III trial of prophylactic cranial irradiation compared with observation in patients with locally advanced non-small-cell lung cancer: Neurocognitive and quality-of-life analysis,” J. Clin. Oncol., vol. 29, no. 3, pp. 279–286, 2011.\\

\noindent[24]	M. B. Pulsifer, R. V Sethi, K. A. Kuhlthau, S. M. MacDonald, N. J. Tarbell, and T. I. Yock, “Early Cognitive Outcomes Following Proton Radiation in Pediatric Patients With Brain and Central Nervous System Tumors,” Int. J. Radiat. Oncol. Biol. Phys., vol. 93, no. 2, pp. 400–407, Oct. 2015.\\

\noindent[25]	R. C. Malenka, J. A. Kauer, R. S. Zucker, and R. A. Nicoll, “Postsynaptic calcium is sufficient for potentiation of hippocampal synaptic transmission,” Science (80-. )., vol. 242, no. 4875, pp. 81–84, 1988.\\

\noindent[26]	A. Chakraborti, A. Allen, B. Allen, S. Rosi, and J. R. Fike, “Cranial irradiation alters dendritic spine density and morphology in the hippocampus,” PLoS One, vol. 7, no. 7, pp. 1–9, 2012.\\

\noindent[27]	I. V. Sokolova, C. J. Schneider, M. Bezaire, I. Soltesz, R. Vlkolinsky, and G. A. Nelson, “Proton radiation alters intrinsic and synaptic properties of CA1 pyramidal neurons of the mouse hippocampus,” Radiat. Res., vol. 183, no. 2, pp. 208–218, 2015.\\

\noindent[28]	M. MacHida, G. Lonart, and R. A. Britten, “Low (60 cGy) doses of 56Fe HZE-particle radiation lead to a persistent reduction in the glutamatergic readily releasable pool in rat hippocampal synaptosomes,” Radiat. Res., vol. 174, no. 5, pp. 618–623, 2010.\\

\noindent[29]	L. Bayarchimeg, M. Batmunkh, A. N. Bugay, and O. Lkhagva, “Evaluation of Radiation-Induced Damage in Membrane Ion Channels and Synaptic Receptors,” Phys. Part. Nucl. Lett., vol. 16, no. 1, pp. 54–62, 2019.\\

\noindent[30]	O. V. Belov, M. Batmunkh, S. Incerti, and O. Lkhagva, “Radiation damage to neuronal cells: Simulating the energy deposition and water radiolysis in a small neural network,” Phys. Medica, vol. 32, no. 12, pp. 1510–1520, 2016.\\

\noindent[31]	M. Batmunkh, L. Bayarchimeg, A. N. Bugay, and O. Lkhagva, “Computer simulation of radiation damage mechanisms in the structure of brain cells,” AIP Conf. Proc., vol. 2377, p. 050001, 2021.\\

\noindent[32]	M. Alp, V. K. Parihar, C. L. Limoli, and F. A. Cucinotta, “Irradiation of Neurons with High-Energy Charged Particles: An In Silico Modeling Approach,” PLoS Comput. Biol., vol. 11, no. 8, pp. 1–21, 2015.\\

\noindent[33]	G. A. Ascoli, D. E. Donohue, and M. Halavi, “NeuroMorpho.Org: A central resource for neuronal morphologies,” J. Neurosci., vol. 27, no. 35, pp. 9247–9251, 2007.\\

\noindent[34]	S. H. Lee et al., “Parvalbumin-positive basket cells differentiate among hippocampal pyramidal cells,” Neuron, vol. 82, no. 5, pp. 1129–44, 2014.\\

\noindent[35]	K. M. Harris, F. E. Jensen, and B. Tsao, “Three-dimensional structure of dendritic spines and synapses in rat hippocampus (CA1) at postnatal day 15 and adult ages: implications for the maturation of synaptic physiology and long-term potentiation,” J. Neurosci., vol. 12, no. 7, pp. 2685–2705, Jul. 1992.\\

\noindent[36]	Y. Loewenstein, A. Kuras, and S. Rumpel, “Multiplicative Dynamics Underlie the Emergence of the Log-Normal Distribution of Spine Sizes in the Neocortex In Vivo,” J. Neurosci., vol. 31, no. 26, pp. 9481–9488, Jun. 2011.\\

\noindent[37]	J. Perl, J. Shin, J. Schümann, B. Faddegon, and H. Paganetti, “TOPAS: An innovative proton Monte Carlo platform for research and clinical applications,” Med. Phys., vol. 39, no. 11, pp. 6818–37, 2012.\\

\noindent[38]	B. Faddegon et al., “The TOPAS tool for particle simulation, a Monte Carlo simulation tool for physics, biology and clinical research,” Phys. Medica, vol. 72, pp. 114–121, 2020.\\

\noindent[39]	J. Allison et al., “Recent developments in GEANT4,” Nucl. Instruments Methods Phys. Res. Sect. A Accel. Spectrometers, Detect. Assoc. Equip., vol. 835, pp. 186–225, 2016.\\

\noindent[40]	S. Agostinelli et al., “GEANT4 - A simulation toolkit,” Nucl. Instruments Methods Phys. Res. Sect. A Accel. Spectrometers, Detect. Assoc. Equip., vol. 506, no. 3, pp. 250–303, 2003.\\

\noindent[41]	J. Allison et al., “Geant4 developments and applications,” IEEE Trans. Nucl. Sci., vol. 53, no. 1, pp. 270–278, 2006.\\

\noindent[42]	J. Schuemann et al., “TOPAS-nBio: An Extension to the TOPAS Simulation Toolkit for Cellular and Sub-cellular Radiobiology,” Radiat. Res., vol. 191, no. 2, pp. 125–138, 2018.\\

\noindent[43]	A. L. McNamara et al., “Geometrical structures for radiation biology research as implemented in the TOPAS-nBio toolkit,” Phys. Med. Biol., vol. 63, p. 175018, 2018.\\

\noindent[44]	S. Incerti et al., “Geant4-DNA example applications for track structure simulations in liquid water: A report from the Geant4-DNA Project,” Med. Phys., vol. 45, no. 8, pp. e722–e739, 2018.\\

\noindent[45]	M. A. Bernal et al., “Track structure modeling in liquid water: A review of the Geant4-DNA very low energy extension of the Geant4 Monte Carlo simulation toolkit,” Phys. Medica, vol. 31, no. 8, pp. 861–874, 2015.\\

\noindent[46]	S. Incerti et al., “Comparison of GEANT4 very low energy cross section models with experimental data in water,” Med. Phys., vol. 37, no. 9, pp. 4692–708, 2010.\\

\noindent[47]	S. Incerti et al., “The Geant4-DNA project,” Int. J. Model. Simulation, Sci. Comput., vol. 1, no. 2, pp. 157–178, 2010.\\

\noindent[48]	J. Ramos-Méndez, J. Perl, J. Schuemann, A. McNamara, H. Paganetti, and B. Faddegon, “Monte Carlo simulation of chemistry following radiolysis with TOPAS-nBio,” Phys. Med. Biol., vol. 63, p. 105014, 2018.\\

\noindent[49]	V. K. Parihar and C. L. Limoli, “Cranial irradiation compromises neuronal architecture in the hippocampus,” Proc. Natl. Acad. Sci. U. S. A., vol. 110, no. 31, pp. 12822–12827, 2013.\\

\noindent[50]	M. Alp and F. A. Cucinotta, “Biophysics Model of Heavy-Ion Degradation of Neuron Morphology in Mouse Hippocampal Granular Cell Layer Neurons,” Radiat. Res., vol. 189, no. 3, pp. 312–325, 2018.\\

\noindent[51]	E. Cacao, V. K. Parihar, C. L. Limoli, and F. A. Cucinotta, “Stochastic Modeling of Radiation-induced Dendritic Damage on in silico Mouse Hippocampal Neurons,” Sci. Rep., vol. 8, no. 1, pp. 1–13, 2018.\\

\noindent[52]	P. M. Klein et al., “Detrimental impacts of mixed-ion radiation on nervous system function,” Neurobiol. Dis., vol. 151, p. 105252, 2021.\\

\noindent[53]	J. Raber et al., “Effects of Six Sequential Charged Particle Beams on Behavioral and Cognitive Performance in B6D2F1 Female and Male Mice,” Front. Physiol., vol. 11, 2020.\\

\noindent[54]	J. Raber et al., “Combined effects of three high-energy charged particle beams important for space flight on brain, behavioral and cognitive endpoints in B6D2F1 female and Male mice,” Front. Physiol., vol. 10, 2019.

\end{document}


\title[Nano-scale simulation of neuronal damage by galactic cosmic rays]{Supplementary Material: Nano-scale simulation of neuronal damage by galactic cosmic rays}

\author{Jonah S Peter$^1$, Jan Schuemann$^2$, Kathryn D Held$^2$, and Aimee McNamara$^2$}

\address{$^1$ Biophysics Program, Harvard University, Boston, MA 02115, United States of America}
\address{$^2$ Department of Radiation Oncology, Massachusetts General Hospital, Boston, MA 02114, United States of America}
\ead{jonahpeter@g.harvard.edu}
\vspace{10pt}
\begin{indented}
\item[]February 14, 2022
\end{indented}

\maketitle 

\appendix
\setcounter{section}{19}

\afterpage{ 
\begin{table}
\caption{\label{S1} Dendritic lengths, neuronal surface areas, and neuronal volumes for each CA1 pyramidal neuron in the Soltesz archive on the NeuroMorpho repository.}
\lineup
\footnotesize
\begin{indented}
\item[]\begin{tabular}{@{}llll}
\br
NMO ID& Dendritic Length ($\mu$m) &Surface Area ($\mu$m$^2$) &Volume ($\mu$m$^3$)\\
\mr
NMO\_36611$^\dagger$& 4391.4&19821.6& 7316.8\\
NMO\_36612&7144.58& 27023.2& 8857.4\\
NMO\_36613&3575.99&	13034.8&	4070.75\\
NMO\_36614&6400.76&	20201.3&	5312.35\\
NMO\_36615&6516.75&	27778.8&	9653.47\\
NMO\_36622&4150.12&	11530.6&	2625.67\\
NMO\_36623&5355.86&	22687.4&	7854.22\\
NMO\_36617&5068.55&	17333.9&	4912.88\\
NMO\_36616&5516.6&	17415.3&	4458.43\\
NMO\_36618&5302.24&	21522.3&	7531.12\\
NMO\_36619&3793.12&	11037.2&	2876.37\\
NMO\_36620&5497.84&	25157&	9498.09\\
NMO\_36621&5358.5&	20066.8&	6268.42\\
\mr
Mean&	5236.3&	19585.4&	6248.9\\
Standard Deviation&	1063.8&	\05463.3&	2405.3\\
\br
\end{tabular}\\
\item[] $^{\dagger}$Indicates the geometry used in this work.
\end{indented}
\end{table}
\clearpage 
}
\normalsize

\afterpage{
\begin{figure}
    \includegraphics[width=\textwidth]{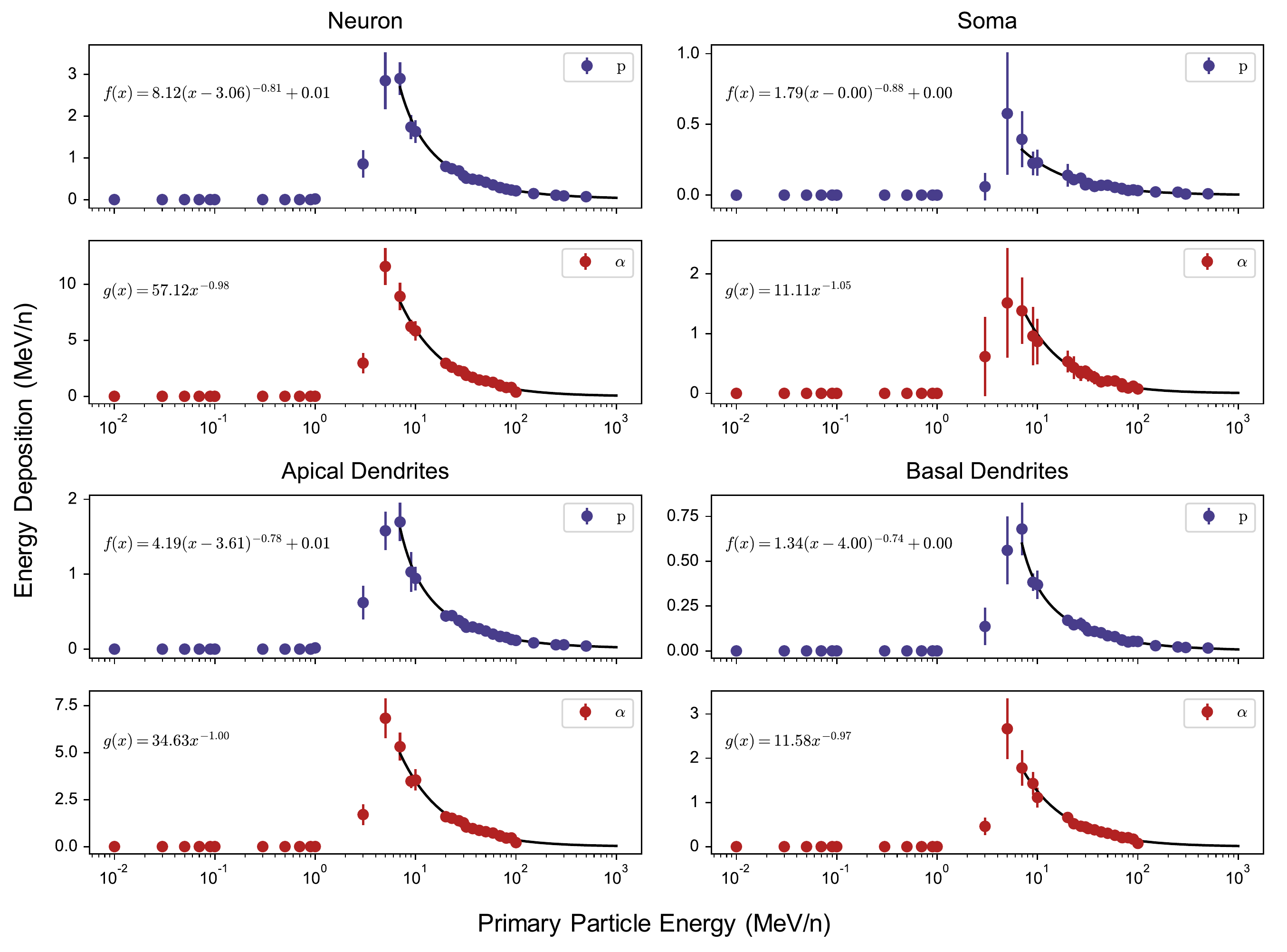}
    \caption{Simulated energy deposition in the neuron and subcellular structures from primary proton (p; navy) and $\alpha$-particle ($\alpha$; red) beams as a function of primary particle energy. Subplots indicate energy deposition in the neuron (top left), soma (top right), apical dendrites (bottom left), and basal dendrites (bottom right). Data points represent mean $\pm$ standard deviation as discussed in the main text. Black curves indicate curve-fitting with the equations shown.}
    \label{fig:S1}
\end{figure}
\clearpage
}

\afterpage{
\begin{figure}
    \includegraphics[width=\textwidth]{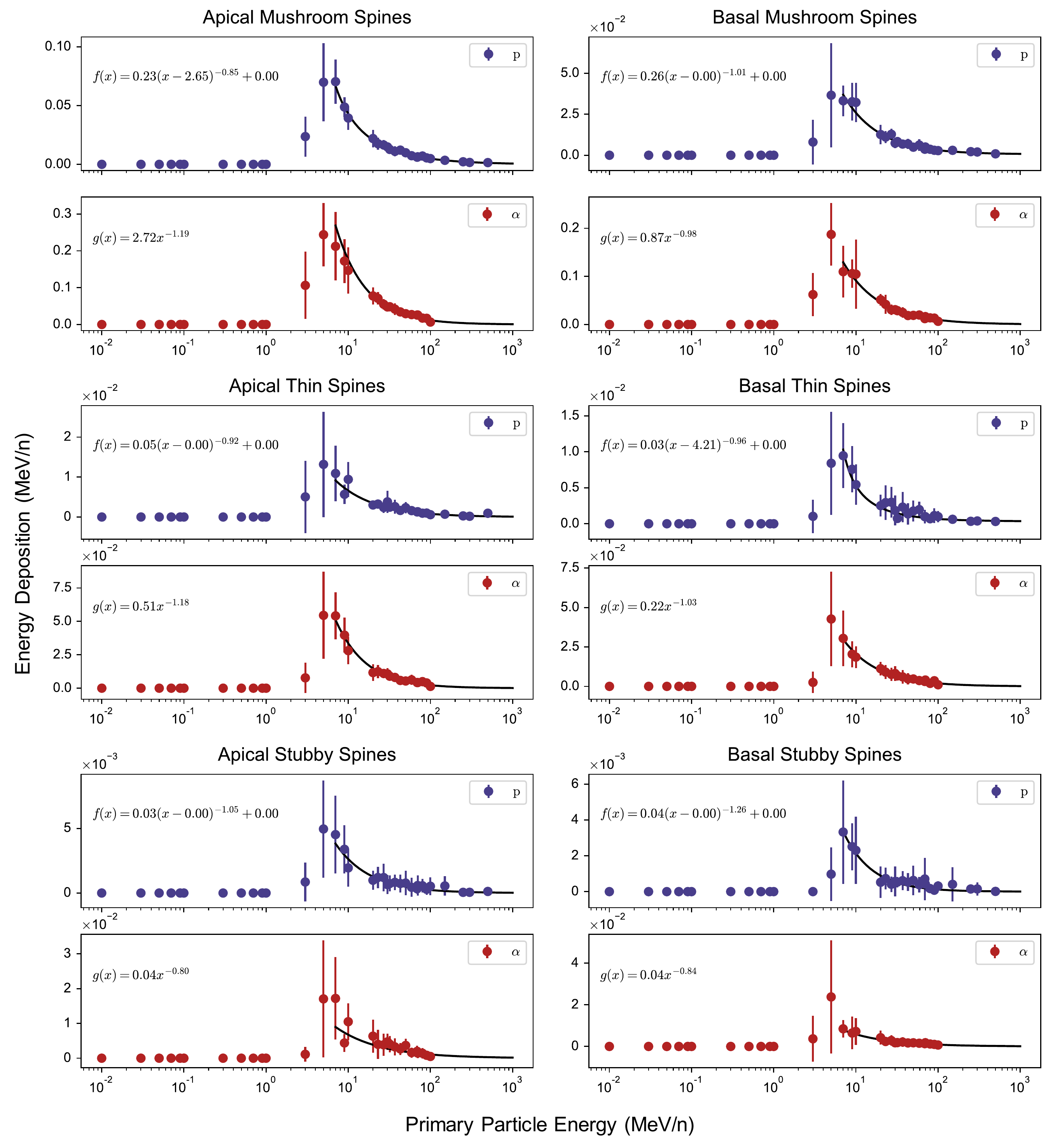}
    \caption{Simulated energy deposition in dendritic spines from primary proton (p; navy) and $\alpha$-particle ($\alpha$; red) beams as a function of primary particle energy. Subplots indicate energy deposition in apical mushroom spines (top left), apical thin spines (middle left), apical stubby spines (bottom left), basal mushroom spines (top right), basal thin spines (middle right), and basal stubby spines (bottom right). Data points represent mean $\pm$ standard deviation as discussed in the main text. Black curves indicate curve-fitting with the equations shown.}
    \label{fig:S2}
\end{figure}
\clearpage
}

\afterpage{
\begin{figure}
    \includegraphics[width=\textwidth]{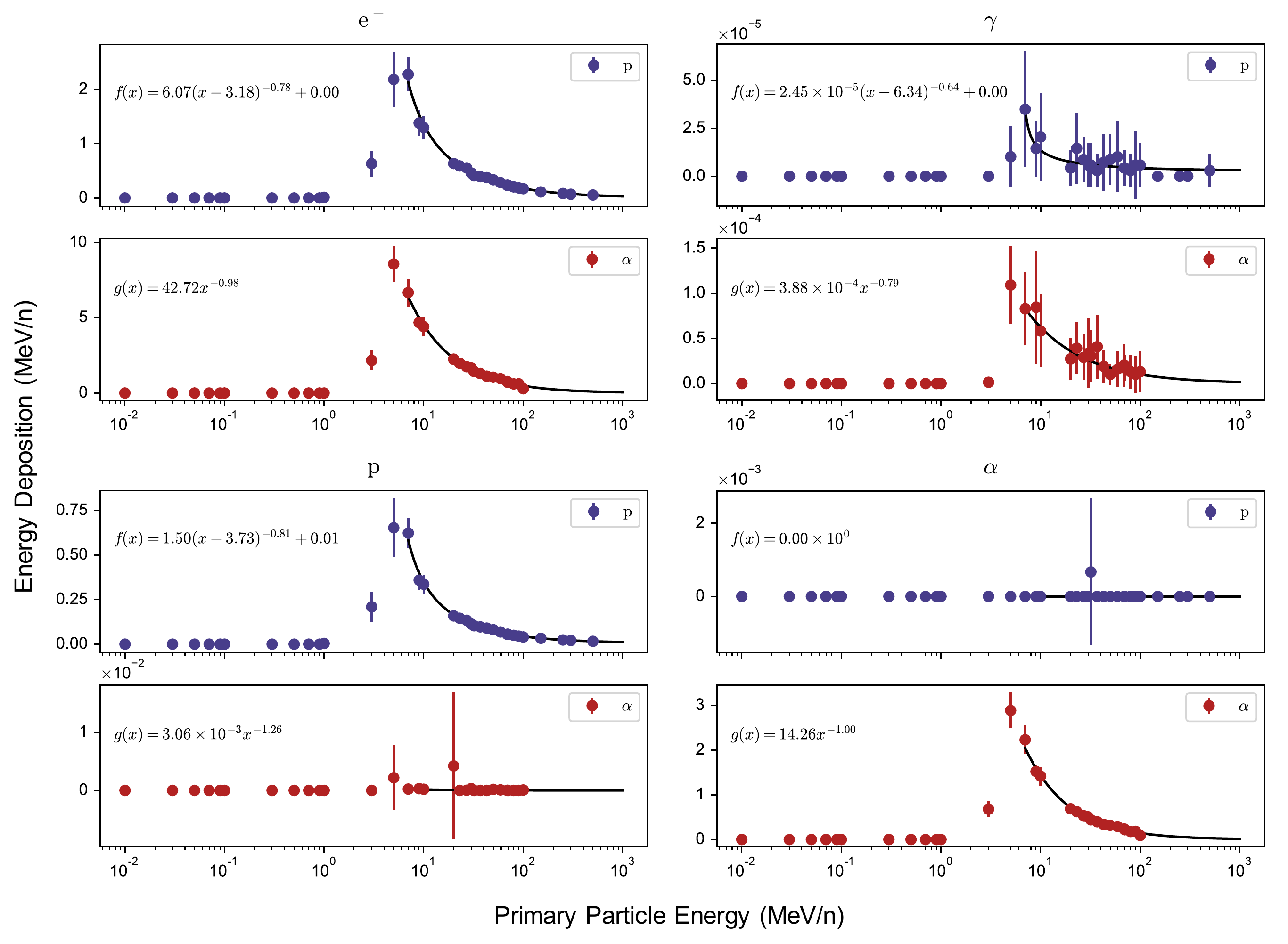}
    \caption{Simulated neuronal energy deposition by major particle species as a function of primary particle energy. Colors indicate the primary ion beam as either proton (p; navy) or $\alpha$-particle ($\alpha$; red) while subplot headers indicate the particle species mediating the energy deposition. The particle species shown are: electron (e$^-$; top left), $\gamma$-radiation ($\gamma$; top right), proton (p; bottom left) and $\alpha$-particle ($\alpha$; bottom right). Minor contributions from additional hadronic species are not shown. Data points represent mean $\pm$ standard deviation as discussed in the main text. Black curves indicate curve-fitting with the equations shown.}
    \label{fig:S3}
\end{figure}
\clearpage
}

\afterpage{
\begin{figure}
    \includegraphics[width=\textwidth]{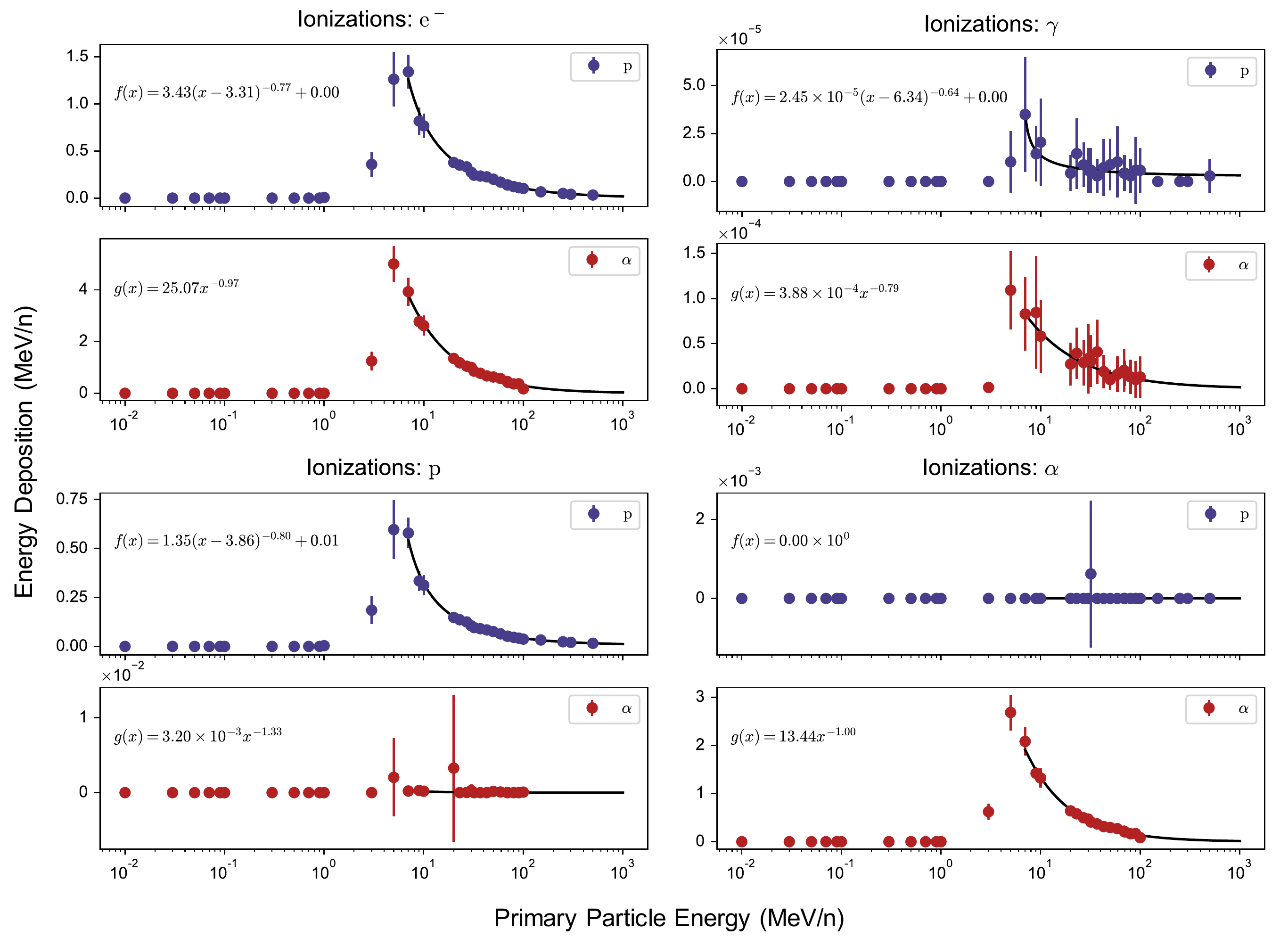}
    \caption{Simulated neuronal energy deposition from ionizations by major particle species as a function of primary particle energy. Colors indicate the primary ion beam as either proton (p; navy) or $\alpha$-particle ($\alpha$; red) while subplot headers indicate the particle species mediating the ionizations. The particle species shown are: electron (e$^-$; top left), $\gamma$-radiation ($\gamma$; top right), proton (p; bottom left) and $\alpha$-particle ($\alpha$; bottom right). Minor contributions from additional hadronic species are not shown. Data points represent mean $\pm$ standard deviation as discussed in the main text. Black curves indicate curve-fitting with the equations shown. }
    \label{fig:S4}
\end{figure}
\clearpage
}

\afterpage{
\begin{figure}
    \includegraphics[width=\textwidth]{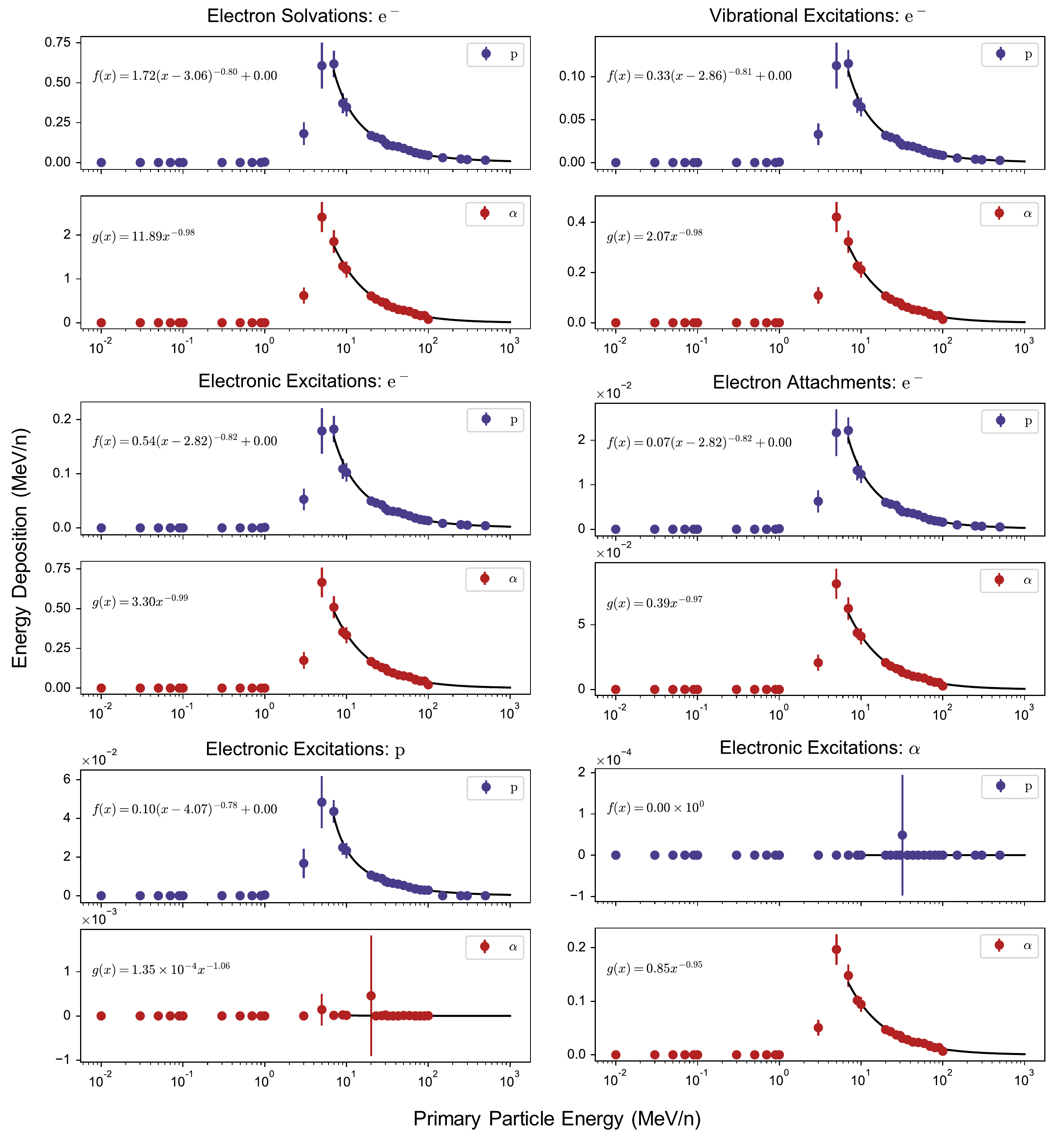}
    \caption{Simulated neuronal energy deposition from additional physics processes as a function of primary particle energy. Colors indicate the primary ion beam as either proton (p; navy) or $\alpha$-particle ($\alpha$; red) while subplot headers indicate the physics process and particle species mediating the energy deposition. The processes shown are: electron solvations (top left), electron-mediated vibrational excitations (top right), electron-mediated electronic excitations (middle left), electron attachments (middle right), proton-mediated electronic excitations (bottom left), and $\alpha$-particle-mediated electronic excitations (bottom right). Minor contributions from additional physics processes and hadronic species are not shown. Data points represent mean $\pm$ standard deviation as discussed in the main text. Black curves indicate curve-fitting with the equations shown.}
    \label{fig:S5}
\end{figure}
\clearpage
}

\afterpage{
\begin{figure}
    \includegraphics[width=\textwidth]{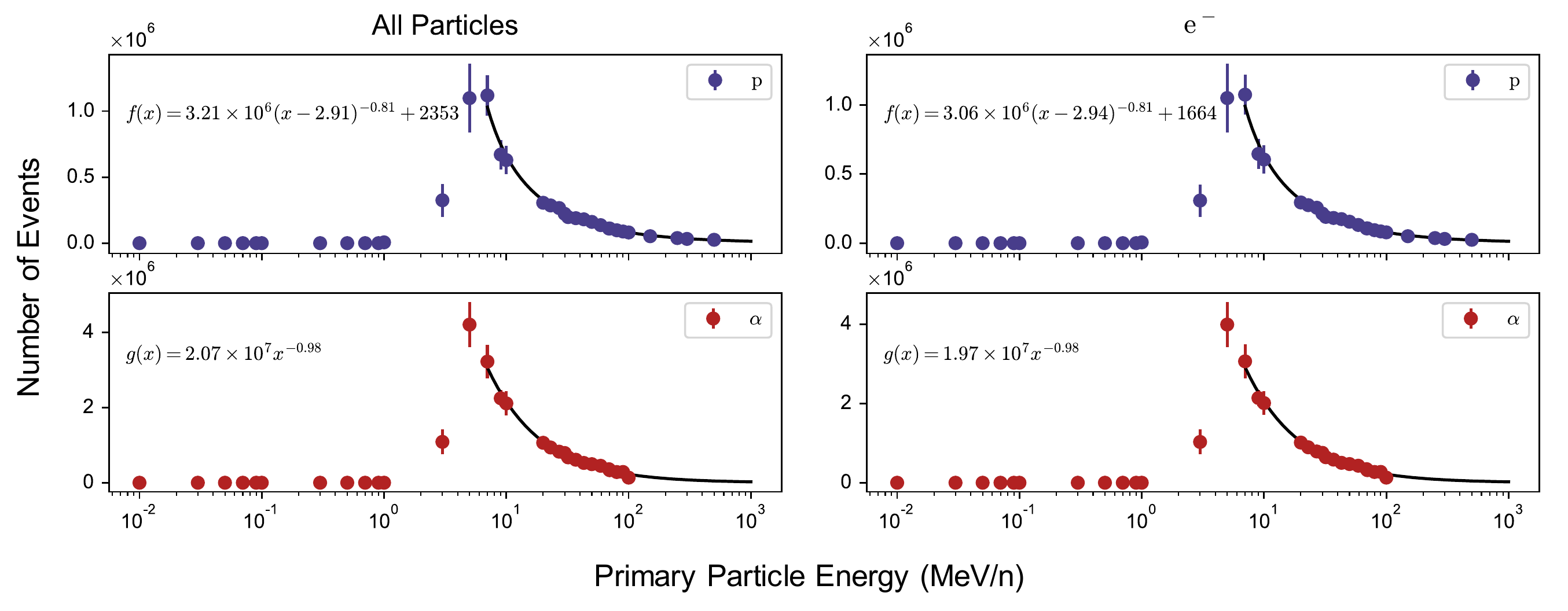}
    \caption{Simulated frequency of neuronal energy deposition events from all particles (left) and secondary electrons (right) as a function of primary particle energy. Colors indicate the primary ion beam as either proton (p; navy) or $\alpha$-particle ($\alpha$; red). Data points represent mean $\pm$ standard deviation as discussed in the main text. Black curves indicate curve-fitting with the equations shown.}
    \label{fig:S6}
\end{figure}
\clearpage
}

\afterpage{
\begin{figure}
    \includegraphics[width=\textwidth]{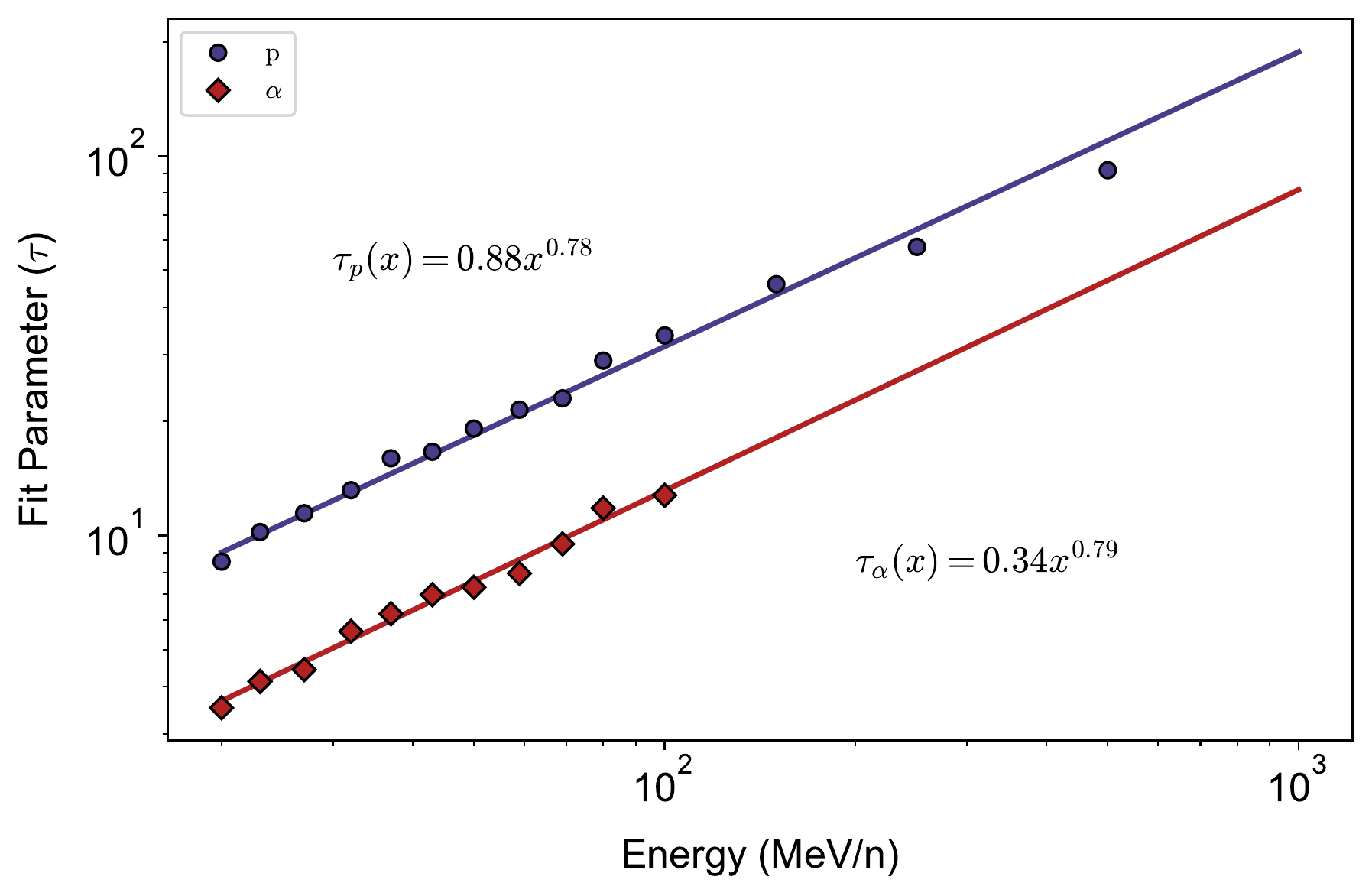}
    \caption{Values for the dendritic hit curve fit parameter $\tau_k(x)$ for protons (p; navy circles) and $\alpha$-particles ($\alpha$; red diamonds) as a function of primary particle energy. Error bars due to curve-fitting uncertainty are smaller than the marker size. Solid lines indicate power-law fits according to the equations shown. These equations were used to extrapolate the values of $\tau_k(x)$ at higher energies.}
    \label{fig:S7}
\end{figure}
\clearpage
}

\afterpage{
\begin{table}
\caption{\label{S2} Values of the characteristic dose parameter $1/\tau_{ij}$ for each ion-energy beam in the GCRSim spectrum. $^{48}$Ti was substituted with $^{56}$Fe as discussed in the main text. Uncertainties due to variability in the curve-fitting process are significantly smaller than anticipated statistical fluctuations and are therefore omitted. (p: proton; $\alpha$: $\alpha$-particle).}
\lineup
\footnotesize
\begin{indented}
\item[]\begin{tabular}{@{}lll}
\br
Ion & Energy (MeV/n) & $1/\tau_{ij}$\\
\mr
p & \0\020 & 0.12\\
p & \0\023 & 0.098\\
p & \0\027 & 0.087\\
p & \0\032 & 0.076\\
p & \0\037 & 0.063\\
p & \0\043 & 0.060\\
p & \0\050 & 0.052\\
p & \0\059 & 0.047\\
p & \0\069 & 0.044\\
p & \0\080 & 0.035\\
p & \0100 & 0.030\\
p & \0150 & 0.022\\
p & \0250 & 0.017\\
p & 1000 & 0.0053$^{\dagger}$\\
$\alpha$ & \0\020 & 0.28\\
$\alpha$ & \0\023 & 0.24\\
$\alpha$ & \0\027 & 0.23\\
$\alpha$ & \0\032 & 0.18\\
$\alpha$ & \0\037 & 0.16\\
$\alpha$ & \0\043 & 0.14\\
$\alpha$ & \0\050 & 0.14\\
$\alpha$ & \0\059 & 0.12\\
$\alpha$ & \0\069 & 0.11\\
$\alpha$ & \0\080 & 0.085\\
$\alpha$ & \0100 & 0.078\\
$\alpha$ & \0150 & 0.055$^{\dagger}$\\
$\alpha$ & \0250 & 0.037$^{\dagger}$\\
$\alpha$ & 1000 & 0.012$^{\dagger}$\\
$^{12}$C & 1000 & 0.16\\
$^{16}$O & \0350 & 0.39\\
$^{28}$Si & \0600 & 0.52\\
$^{56}$Fe & \0600 & 1.3\\
$^{48}$Ti ($^{56}$Fe) & 1000 & 0.76\\
\br
\end{tabular}\\
\item[] $^{\dagger}$Denotes extrapolated value.
\end{indented}
\end{table}
\clearpage
}
\normalsize

\afterpage{
\begin{table}
\caption{\label{S3} The \% contribution of each physics process to the neuronal energy deposition and number of energy deposition events for each of the two fluence distributions. Values are mean $\pm$ standard deviation as discussed in the main text.}
\footnotesize
\lineup
\begin{indented}
\item[]\begin{tabular}{@{}lllll}
\br
&\centre{2}{Energy Deposition (\%)} & \centre{2}{Number of Events (\%)}\\
\ns
\ns
&\crule{2} &\crule{2}\\
Physics Process & GCRSim & SimGCRSim & GCRSim & SimGCRSim\\
\mr
Ionizations&$68\m\0\pm 1$&$69\m\0\pm 2$&$14\m\0\pm 1$&$15\m\0\pm 1$\\
Electron Solvations&$21\m\0\pm 1$&$21\m\0\pm 2$&$14\m\0\pm 1$&$15\m\0\pm 1$\\
Electronic Excitations&$\06.7\0\pm 0.5$&$\06.2\0\pm 0.7$&$\01.8\0\pm 0.2$&$\01.7\0\pm 0.2$\\
Vibrational Excitations&$\03.6\0\pm 0.2$&$\03.5\0\pm 0.3$&$70\m\0\pm 2$&$69\m\0\pm 2$\\
Electron Attachments&$\00.72\pm 0.05$&$\00.71\pm 0.07$&$\00.23\pm 0.02$&$\00.23\pm 0.03$\\
\br
\end{tabular}
\end{indented}
\end{table}
\clearpage
}
\normalsize

\afterpage{
\begin{table}
\caption{\label{S4} The number of energy deposition events from all physics processes and ionizations only in various subcellular volumes for each of the two fluence distributions. Values are mean $\pm$ standard deviation and are given per 0.5 Gy of neuronal absorbed dose, as discussed in the main text.}
\lineup
\footnotesize
\begin{indented}
\item[]\begin{tabular}{@{}lllll}
\br
&\centre{2}{All Processes} & \centre{2}{Ionizations}\\
\ns
\ns
&\crule{2} &\crule{2}\\
Volume Compartment & GCRSim & SimGCRSim & GCRSim & SimGCRSim\\
\mr
Events per micron of dendrite&$1760\pm 1$&$1700\pm 200$&$250\pm 10$&$250\pm 10$\\
Events per spine&$\0130 \pm 10$&$\0120 \pm 20$&$\019 \pm 2$&$\018 \pm 3$\\
Events per mushroom spine&$\0330 \pm 80$&$\0310 \pm 90$&$\050 \pm 10$&$\050 \pm 10$\\
Events per thin spine&$\0\050 \pm 20$&$\0\050 \pm 20$&$\0\07 \pm 2$&$\0\07 \pm 3$\\
Events per stubby spine&$\0\030 \pm 10$&$\0\020 \pm 10$&$\0\04 \pm 2$&\0\0$3 \pm 1$\\
\br
\end{tabular}
\end{indented}
\end{table}
\clearpage
}
\normalsize